\documentclass[useAMS,usenatbib]{mn2e}
\usepackage{graphicx}
 \usepackage{deluxetable}

\newcommand{\CH}{\hbox{[C/H]}}
\newcommand{\SiH}{\hbox{[Si/H]}}

\newcommand{\CIV}{\hbox{{\rm C}{\sc \,iv}}}
\newcommand{\OVI}{\hbox{{\rm O}{\sc \,vi}}}

\newcommand{\SiIV}{\hbox{{\rm Si}{\sc \,iv}}}

\newcommand{\HI}{\hbox{{\rm H}{\sc \,i}}}

\newcommand{\Htwo}{\hbox{{\rm H}$_2$}}
\newcommand{\HeI}{\hbox{{\rm He}{\sc \,i}}}
\newcommand{\lya}{\hbox{{\rm Ly}$\alpha$}}

\newcommand{\Ha}{\hbox{{\rm H}$\alpha$}}

\newcommand{\mpy}{\hbox{M$_{\odot}$\,yr$^{-1}$}}

\newcommand{\msun}{\hbox{M$_{\odot}$}}
\newcommand{\rhostarsun}{\hbox{M$_{\odot}$\,yr$^{-1}$\, Mpc$^{-3}$}}
\newcommand{\rhosun}{\hbox{M$_{\odot}$\, Mpc$^{-3}$}}
\newcommand{\Lsun}{\hbox{L$_{\odot}$\, Mpc$^{-3}$}}
\newcommand{\cmsq}{\hbox{cm$^{-2}$}}
\newcommand{\NHI}{\hbox{$N_{\HI}$}}

\newcommand{\kms}{\hbox{${\rm km\,s}^{-1}$}}
\newcommand{\nn}{\nonumber}
\newcommand{\LZ}{\hbox{$L$--$Z$}}
\newcommand{\rhoZlocal}{\hbox{$5.0\times 10^6$~\rhosun}} 
\newcommand{\rhoOlocal}{\hbox{$3.0\times 10^6$~\rhosun}}

\def\bsp_small{\vspace{0.5cm}\small\noindent This paper
has been typeset from a \TeX / \LaTeX\ file prepared by the author.}

\title[The missing metals  problem. III]{The missing metals  problem. III How many metals are expelled from galaxies?}
\author[N. Bouch\'e et al.]{
 \parbox[t]{\textwidth}{
  Nicolas Bouch\'e$^1$\thanks{E-mail: nbouche@mpe.mpg.de (NB)},
  Matthew D. Lehnert$^{1,2}$,
  Anthony Aguirre$^3$,
  C\'eline P\'eroux$^4$, 
  Jacqueline Bergeron$^5$
 }
 \vspace*{0.6cm}\\
  $^1$Max Planck Institut f\"ur extraterrestrische Physik, Giessenbachstra\ss e, D-85748 Garching, Germany \\
  $^2$Laboratoire Galaxies Etoiles Physique et Instrumentation (GEPI),
l'Observatoire de Paris, 5 place Jules Janssen, 92195 Meudon France \\
  $^3$University of California, Santa Cruz, CA\\
  $^4$European Southern Observatory, Karl-Schwarzschild-Str 2, D-85748 Garching, Germany \\ 
  $^5$Institut d'Astrophysique de Paris, Paris, France \\
  \vspace{-0.5cm}
}
\begin{document}

\date{Accepted 14 March 2007}

\pagerange{\pageref{firstpage}--\pageref{lastpage}}
%\pubyear{2004}

\maketitle

\label{firstpage}

%%%%%%%%%%%%%%%%%%%%  - ABSTRACT  %%%%%%%%%%%%%%%%%%%%%%%
\begin{abstract}
We revisit the metal budget at $z\simeq 2$, and include the contribution of
the  intergalactic medium. Past estimates of the metal budget indicated that, at redshift $z\simeq2.5$,  90\%\ of the expected metals were missing.
In the first two papers of this series, we already showed that
 $\sim$30\%\ of the metals are observed in all $z\sim2.5$ galaxies   detected in current surveys.
This fraction  could increase to $\la$ 60\%\ if one extrapolates the faint end of the LF, leaving $>$40\%\
of the metals missing.
Here, we extend our analysis to the metals outside galaxies, i.e. in 
intergalactic medium (IGM), using (1) observational data and (2) analytical calculations.
Our results for the two   are strikingly similar:
(1) Observationally, we  find that, besides the small (5\%)
contribution of DLAs, the forest and sub-DLAs contribute subtantially   to make $\la$30--45\%\ of the metal budget,
but neither of these appear to be sufficient to close the metal budget.
The forest accounts for 15--30\%\ depending on the UV background, and sub-DLAs for $\ga$2\% to $\la$17\%\ depending on the
ionization fraction. Combining the metals in galaxies and in the IGM,
  it  appears now that  $>$65\%\ of the metals have been accounted for, 
and the `missing metals' problem is substantially eased.
(2) We perform analytical calculations based  on  the effective yield--mass ($y_{\rm eff}$--$V_c$) relation,
whose deficit for small galaxies  is   considered   as evidence for supernova driven outflows.
As a test of the method, we show that, at $z=0$, the calculation self-consistently predicts the total amount of metals   expelled from galaxies.
At $z=2$, we find  that the method predicts that  $25$--$50$\%\ of the metals  
have been ejected from galaxies into the IGM, consistent with the observations ($\la$35\%).
The metal ejection is predominantly by $L_B<\frac{1}{3}L_B^*(z=2)$  galaxies, 
which are responsible for 90\%\  the metal enrichment,
while the 50 percentile is at $L\sim \frac{1}{10}L^*_B(z=2)$. 
As a consequence, if indeed $50$\%\ of the metals  
have been ejected from galaxies, 3--5 bursts of star formation are required per galaxy
prior to $z=2$.
The ratio between the mass of metals outside galaxies to those in stars has changed  from $z=2$ to $z=0$:
it was 2:1 or 1:1 and is now 1:8 or 1:9. This evolution implies that
a significant fraction of the IGM metals will cool and fall back into galaxies.
\end{abstract}


%%%%%%%%%%%%%%%%%%%%%%%%% INTRODUCTION %%%%%%%%%%%%%%%%%%%%%%%
\section{Introduction}
\label{section:intro}

Locally, our picture of the metal (and baryon) budget has become more and more complete
over the past few years \citep{FukugitaM_98a,FukugitaM_04a}. Roughly speaking, 30\%\ of the baryons are
in the \lya\ forest \citep{StockeJ_04a}, 50\%\ are in a warm-hot phase \citep[WHIM,][]{TrippT_04a}, 5-10\%\ are in the 
intra-cluster medium (ICM), and 10\%\ are in stars.
So, even-though WHIM \citep[e.g.][]{SembachK_04a,NicastroF_05a} and intra-group medium \citep{FukugitaM_04a}
dominate the baryon budget, they    contain a minor fraction ($<10$~\%) of
the metals. 
Indeed, about 80--90\%\ of all the metals produced by type~II and type~Ia supernovae are locked in stars (10\%) and stellar remnants (80\%)
such as white dwarfs, neutron stars and black holes \citep{FukugitaM_04a}.

Ten Gyrs ago ($z\simeq 2.5$), the situation was very different. Most (90\%) of the baryons
were in the \lya\ forest \citep[e.g.][]{RauchM_97a,PentonS_00a,SchayeJ_01b,SimcoeR_04a}, but
 our knowledge of metal abundances is still highly incomplete.
 In the past, it was realized that only a small fraction  (20\%) of the expected metals
  is accounted for when one adds the contribution
of the \lya\ forest ($N_{\HI}=10^{13-17}$~\cmsq), damped \lya\  absorbers (DLAs) ($N_{\HI}>10^{20.3}$~\cmsq),
 and  galaxies such as Lyman break galaxies (LBGs) \citep{PettiniM_99a,PagelB_02a,PettiniM_03b}. 

As discussed in \citet{PettiniM_03a}, either
the missing metals were expelled into the intergalactic medium (IGM) via galactic winds \citep[as already discussed by][]{LarsonR_75a}
since LBGs drive   winds \citep{AdelbergerK_03a,ShapleyA_03a} much
like  the one seen locally \citep{LehnertM_96a,DahlemM_97a,LehnertM_99a,HeckmanT_00a,MartinC_99a}, or they are
in a galaxy population not accounted for so far. 

In \citet{BoucheN_05c} (hereafter paper~I),  we  showed that only 5\% (and $\la9$\%) of the expected metals
are in submm selected galaxies (SMGs) and in \citet{BoucheN_06a} (hereafter paper~II),
we revisited the missing metal problem  in light of the several new $z\sim2$ galaxy populations,
such as the `distant red galaxies' (DRGs) and the 'BX' galaxies.
Paper~II showed that the contribution of $z=2$ galaxies
amounts to 18\%\ for the star forming galaxies (including 8\%\ for the rarer $K$-bright galaxies with $K_s<20$)
and to 5\%\ for the DRGs ($J-K>2.3$).
Thus, adding the contribution of star-forming `BX' galaxies, DRGs, and the SMGs,
the total contribution of $z\sim2$ galaxies is at least $18+5+5\sim30$\%\ of the metal budget.
As shown in paper~II, if one extrapolate these results to the faint end of the luminosity function,
$\la$60\%\ of the metals are in $z\sim2$ galaxies, leaving $\ga40$\%\ missing.

The goal of this paper is to  use  observational data (section~\ref{section:metalbudget}) and
 analytical calculations (section~\ref{section:alternative})
to gain insights on  the missing metals.
In section~\ref{section:metalbudget}, we  include the contribution of metals in various gas phases,
 traced   by QSO absorption lines. The dominant
contributors are the forest (with $\log \NHI=10^{13\hbox{--}17}$) and the sub-DLAs with $\log \NHI=10^{19\hbox{--}20.3}$.
We will show that neither the forest alone, nor sub-DLAs, can account for the remaining 40\%\ of missing metals.
In section~\ref{section:alternative}, we  explore the possibility that much of the remaining missing metals were ejected
from galaxies  based on `effective yield' arguments. Namely, locally small galaxies have lost a significant
fraction of their metals,  reflected by their offset from the closed-box expected yield \citep[e.g.][]{PilyuginL_98a,KoppenJ_99a,GarnettD_02a,DalcantonJ_06a}.
Indeed, if all galaxies were evolving as `closed boxes',
their metallicity $Z$ would be inversely proportional to the gas fraction $\mu$, i.e. $Z\;\propto\;-\ln(\mu)$, and 
they would all reach approximately solar metallicity once they convert their gas into stars \citep[e.g.][]{SearleL_72a,AudouzeJ_76a,TinsleyB_78a,EdmundsM_90a}.
The effective yield, $y_{\rm eff}$, which is simply the proportionality constant,
is defined as:
\begin{eqnarray}
y_{\rm eff} &=& \frac{Z}{\ln(1/\mu)} \label{eq:effective:yield}\;.
\end{eqnarray}
The ratio between the solar yield and $y_{\rm eff}$ gives the fraction of metals
that were ejected (see section~3.1).

In the remainder of this paper, we used a $H_0=70~h_{70}$~\kms~Mpc$^{-1}$, $\Omega_M=0.3$ and $\Omega_\Lambda=0.7$.
We use a $Z_\odot=0.0189$ and a Salpeter IMF throughout.

\section{The metal budget}
\label{section:metalbudget}

\subsection{The Metal Production Rate and Assumptions}

Stellar nucleosynthesis and star formation   govern the production of metals.
Some of the metals remain locked up in stellar remnants or long-lived stars, some are returned  
into the interstellar medium (ISM) and another fraction is expelled into the intergalactic medium (IGM) via galactic winds and outflows.
Under the assumption of instantaneous recycling approximation~\footnote{The instantaneous recycling approximation applies here
 since the timescale for massive stellar evolution ($10^7$~yr for a $M\geq 15$~\msun\ star)
    is much shorter than cosmic timescales ($10^9$ yr) at $z\simeq  2.5$.} \citep{SearleL_72a}, the global  metal
production rate $\dot{\rho}_Z$ can be directly related to the star formation rate density (SFRD) $\dot{\rho}_{\star}$
 through the IMF-weighted yield $<p_z>$ \citep{SearleL_72a,SongailaA_90a,MadauP_96a}:
\begin{eqnarray}
\dot{\rho}_{Z}&=&<p_z>\;\dot{\rho}_{\star}\;, \hbox{with}\label{eq:dotrhoZ} \\
<p_z>&=&\int \mathrm d m \; p_z(m) m \phi(m) \;, \label{eq:meanyield}
\end{eqnarray}
where $\phi(m)$ is the IMF, and $p_z(m)$ is the metal yield for stars of mass $m$.
The total expected  amount of metals $\rho_{Z,\rm expected}$ formed by a given time $t$ is the
integral of Eq.~\ref{eq:dotrhoZ} over time.

The   IMF-weighted yield in Eq.~\ref{eq:dotrhoZ}  is equal to $\frac{1}{42}$ or 2.4\%\ \citep{MadauP_96a}
using a Salpeter IMF   and the type~II   stellar yields (for solar metallicity)
$p_z(m)$ from \citet{WoosleyS_95a}~\footnote{The contribution from stellar winds from massive stars is negligeable \citep{HirschiR_05a}.}.
It should be noted that changing the IMF (or the low mass end) will change both $\dot{\rho}_{\star}$ and $p_z$  in the same
way leaving $\rho_Z$ unchanged since $\dot{\rho}_{\star}$ is measured from a stellar luminosity density $\rho_L$.
In other words,  the metal production rate is directly related to  the mean luminosity
density $\rho_L$ via $\rho_L=\epsilon \; \dot{\rho}_Z$, where $\epsilon$ is independent of the IMF \citep{SongailaA_90a,MadauP_96a}.

The SFRD in  Eq.~\ref{eq:dotrhoZ}, $\dot{\rho}_{\star}$, 
is  inferred from rest-frame UV surveys and at redshifts greater than 1, 
 appears to be constant from a redshift of $z\sim4$
to $z\sim1$ at a level of about $\dot{\rho}_{\star}\simeq 0.1\;h_{70}$~\mpy~Mpc$^{-3}$ 
{\it once corrected for dust extinction} \citep{LillyS_96a,MadauP_96a,SteidelC_99a,DickinsonM_03a,GiavaliscoM_04a,DroryN_04a,HopkinsA_04a,HopkinsA_06a,PanterB_06a,FardalM_06a}.

\begin{figure}
%\centerline{\includegraphics[width=80mm]{figs/SFH_history.1.eps}}
\centerline{\includegraphics[width=80mm]{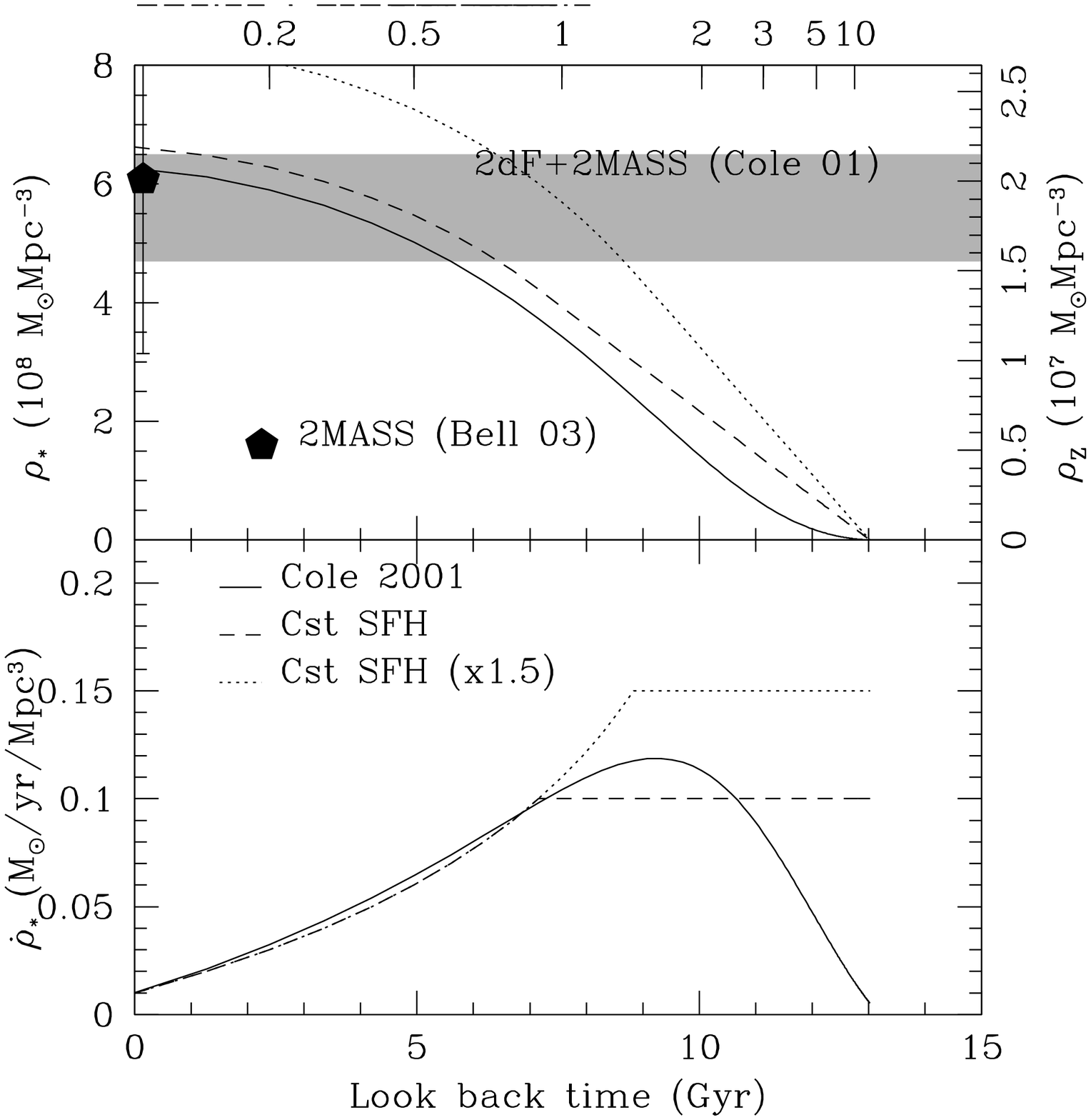}}
\caption{{\it Bottom panel:}  Star formation history  (SFH) of the universe. The solid line shows 
$\dot{\rho}_\star(t)$ parameterized by \citet{ColeS_01a} for a extinction of $E(B-V)=0.10$. 
The dashed line shows $\dot{\rho}_\star(t)$ set to  $0.1\;h_{70}$~\mpy~Mpc$^{-3}$ beyond redshift $z=1$, and linearly
proportional to $z$ below $z=1$. The dotted line shows $\dot{\rho}_\star(t)$ set $0.15\;h_{70}$~\mpy~Mpc$^{-3}$ at high redshifts,
while keeping the same behavior at  low redshifts. {\it Top panel:} Mass stellar build-up (${\rho}_\star(t)$)
for the same three parameterizations. The left axis shows the visible stellar density ($(1-R)\cdot\int$~SFH), where
$R$ is the recycled fraction, while the right axis shows the metal production ($<p_z>\cdot\int$~SFH).
Both the solid and dashed curves reach the observational constraint of
$\rho_\star(z=0)$ \citep{ColeS_01a}, shown by the shaded area. The dotted line clearly overshoots the observations, leaving little
room on the SFR at high-redshifts. As a consequence, the  contributions of LBGs  \citep[dust-corrected,][]{SteidelC_99a} and SMGs \citep{ChapmanS_05a}
cannot simply be added. }
\label{fig:SFH}
\end{figure}

In order to compute $\rho_Z(t)$ from Eq.~\ref{eq:dotrhoZ}, we parameterize the SFRD, $\dot{\rho}_{\star}$,
 as shown in Fig.~\ref{fig:SFH} (bottom).
The solid line shows  $\dot{\rho}_\star(t)$ parameterized by \citet{ColeS_01a} for an extinction of $E(B-V)=0.10$. 
The dashed line shows $\dot{\rho}_\star(t)$ set to  $0.1\;h_{70}$~\mpy~Mpc$^{-3}$ above redshift $z=1$, and linearly
proportional to $z$ below $z=1$. The dotted line shows $\dot{\rho}_\star(t)$ set $0.15\;h_{70}$~\mpy~Mpc$^{-3}$ at high redshifts,
while keeping the same decline below $z\simeq1$. 

We first check that the integrated SFRD gives the amount of  stars 
using $\rho_{\star}=(1-R) \cdot \int_0^t \mathrm d t \;\dot{\rho}_{\star}(t) \;, $
$R$ is the mass fraction recycled into the ISM. For a Salpeter IMF, $R$ equals $0.28$  \citep{ColeS_01a}.
The top panel shows ${\rho}_\star(t)$  for the three parameterizations. 
 Both the solid and dashed curves reach the local stellar mass density
 shown by the shaded area  for   \citep{ColeS_01a} and the filled symbol for \citet{BellE_03a}~\footnote{We converted the results
of \citet{BellE_03a} to a Salpeter IMF by multiplying their results by $\times1.6$.}. 
The agreement was also found  by others \citep[e.g.][]{RudnickG_03a,DickinsonM_03a,RudnickG_06a}. 
Recent studies pointed to some tension between the local $\rho_{\star}(z=0)$ and the integrated SFRD
 \citep{HopkinsA_06a,FardalM_06a}.  

The overlap between different galaxy populations  is very much unknown.
With respect to the SFRD, two populations,    LBGs and SMGs, reaches similar levels  \citep[e.g.][]{SteidelC_99a,ChapmanS_05a},
 after a dust-correction  for LBGs of a factor of $\sim 5$. Should the SFR of these two populations 
  be added, which would double the SFRD at $z=2$?
The dotted line in Fig.~\ref{fig:SFH} shows  that,  if one doubles the SFRD at $z=2$, 
one over-predicts the  observed stellar density $\rho_\star(z=0)$ by $\sim~40$\%\ if
the SFRD is increased by 50\%. 
Thus, there is little room to add the SMG contribution to the SFRD at high-redshifts ($\sim 0.1~\rhostarsun$),
given the large dust-correction applied to LBGs that already encompasses various galaxy populations.
This conclusion is very general and does not depend on the tension between $\rho_\star(z=0)$  and the integral of the SFRD discussed
in \citet{HopkinsA_06a,FardalM_06a}, adding the SMGs, i.e. doubling the SFR at $z\sim2$, would increase the discrepancy further.

\subsection{The expected amount of metals}

The right axis of Fig.\ref{fig:SFH}(top) shows the metal density  $\rho_Z(t)$ found from integrating Eq.~\ref{eq:dotrhoZ} over time.
From this figure, one sees that at $z=0$, the amount of metals formed (by type~II supernovae) is  expected to be~\footnote{In the literature,
 it is sometimes useful to express $\rho_{Z}$ relative to the baryon density 
\citep[e.g.][]{PettiniM_03b,PettiniM_06a}: $\overline Z_b\equiv\rho_Z/\rho_b/Z_\odot$.
  This quantity represents 
the `fraction of baryons with solar metallicity' or the `mean metallicity' of
the universe if the metals were spread uniformly over all the baryons.
 Eq.~\ref{eq:massmetals:z=0} corresponds to
$\overline Z_b 	= 0.20 \frac{<p_z>}{1/42}\frac{0.0189}{Z_\odot}\;h_{70}^{-1}\;\nn,$
which is close to the mean metallicity of the ICM \citep[][and references therein]{RenziniA_04a}.}
\begin{eqnarray}
\rho_{Z, \rm exp.}(z=0)  &=& 2.13\times10^7 \; \frac{<p_z>}{1/42} ~\rhosun \;,\label{eq:massmetals:z=0}
\end{eqnarray}
if we integrate the SFH from $t=13.7$~Gyr until the present.
As a cross-check, this number (Eq.~\ref{eq:massmetals:z=0}) is consistent with 
the findings of \citet{FukugitaM_04a}. Indeed, their Table~3 list
 the amount of metals in different classes of objects.
 Given that the mean yield $<p_z>=1/42$, used in Eq.~\ref{eq:massmetals:z=0}, is calculated including only stars 
 with $m>10$~\msun, we exclude the contribution  from main sequence (MS) stars and from 
 white dwarfs from the table of \citet{FukugitaM_04a}, and
  find  that the   $z=0$ metal density is  $\rho_{Z, \rm observed}(z=0)\sim 2.3\times10^7$~\rhosun, which  is  quite close to Eq.~\ref{eq:massmetals:z=0}. 
  In the remainder of this paper, 
 we will refer to these numbers as `the $z=0$ metal density', but
 one should keep in mind that it does not include the contribution from white dwarfs (SN~Ia) and locked in the MS~(see Table~\ref{table:bigsummary})
 since we will focus on the type~II yields at $z\simeq2.5$ throughout this paper.

At high-redshifts, if we integrate the SFH, 
from $z=4$  to $z=2$, corresponding to a time interval of
    1.68$\;h_{70}^{-1}$~Gyrs, the  expected  amount of metals formed   is 
    \stepcounter{footnote}
\begin{eqnarray}
\rho_{Z, \rm exp.}(z=2)	&\simeq &  4.0\times 10^{6} 
		\frac{\dot{\rho}_{\star}}{0.1}\frac{<p_z>}{1/42}  \;h_{70}^0~\rhosun \;\hbox{or}  \label{eq:massmetals}\\
\overline Z_b &\simeq & 0.035 \frac{<p_z>}{1/42}\frac{0.0189}{Z_\odot}\;h_{70}^{-1}\;\;\hbox{\footnotemark[\value{footnote}]} \nn
\end{eqnarray}
\footnotetext[\value{footnote}]{We note that \citet{PettiniM_06a} uses $\rho_{Z, \rm exp.}\;=\;3.4\;\times\;10^{6}$~\rhosun\ $<p_z>=1/64$,
and $Z_\odot=0.0126$, which gives $Z_b=0.045$.}
By comparing Eq.~\ref{eq:massmetals} with Eq.~\ref{eq:massmetals:z=0}, and from Fig.~\ref{fig:SFH}, one sees 
 that about 1/4 or 1/5  of the metals are already produced by $z=2$ \citep[e.g.][]{PagelB_02a,PettiniM_03b,PettiniM_06a}.
This is consistent with  the stellar mass density evolution studies
\citep[e.g.][]{RudnickG_03a,RudnickG_06a,PanterB_06a} that have found similar amount of evolution.

If we integrate the constant SFH (at 0.1~\mpy~Mpc$^{-3}$) from  $z=10$ to $z=2$, then the amount of metals formed by $z=2$ is
\begin{eqnarray}
\rho_{Z,\rm exp.}(z>2) &\simeq & 6.4\times 10^6 \frac{<p_z>}{1/42} \hbox{$h_{70}^0$~\msun~Mpc$^{-3}$} \label{eq:massmetals:z2}\;.
\end{eqnarray}

\begin{table*}
\caption{The metal budget for $z\sim2$ galaxies.  \label{table:metals:galaxies}}
\begin{tabular}{lllllll}
\hline
Class	&	 $\rho_{Z}$	& 	$\Omega_{Z}$	&	$\overline Z_b$	&  $\rho_Z/\rho_{Z,\rm tot}$~\tablenotemark{a} & Ref~\tablenotemark{b} & Note	\\
	&	($\rhosun$)	&			&	($Z_\odot$)		& (\%)	&	&\\
\hline
SMGs	&	$<3.6\times10^5$	&	$<2.7\times10^{-6}$	&   $<$0.0031	&   $<$9	& 1	&  $>3$mJy ($Z=2\;Z_\odot$) \\
SMGs	&	$1.8\times10^5$	&	$1.6\times10^{-6}$	&   0.0015	&    5  & 1	&  $>3$mJy ($Z=1\;Z_\odot$) \\
BX	&	$3.8\times10^5$ &	$2.8\times10^{-6}$	&   0.0033	&    10 & 2	&  $>L^*$ \\
BX$+K<$20 &	$3.0\times10^5$ &	$2.2\times10^{-6}$	&   0.0027	&    8	& 2   & $>L^*$ \\
DRGs	&	$2.0\times10^5$	&	$1.5\times10^{-6}$	&   0.0018	&    5  & 2	& $J-K>2.3$   \\
\hline
Total Observed	&	$1.1\times10^6$	&	$7.8\times10^{-6}$	&   0.010	&   $\sim$30&  2 & $>L^*$	\\
Total Inferred	&	$2.2\times10^6$	&	$1.6\times10^{-5}$	&   0.020	&   $\sim$60&  2 & all $L$	\\
Missing &			&				&		&  $>40$	&	&  all $L$\\
\hline
\tablenotetext{a}{The fractional contributions are calculated
using the amount of metals expected from the SFH: $\rho_{Z,\rm tot}\;=\;4\times10^6$~\rhosun\ (Eq.~\ref{eq:massmetals}).}
\tablenotetext{b}{References: (1) Paper~I, (2) Paper~II.}
\end{tabular}
\end{table*}

\subsection{Metals in galaxies}

Some argued that the dusty ISM of submm galaxies could harbor most of the remaining missing metals.
However,  in paper~I,  we  showed that the remaining missing metals cannot be
in SMGs based on direct metallicity and gas mass measurements.
Only 5\% (and $\la9$\%) of the expected metals are in SMGs.
 
The second paper of this series (Paper~II) showed that the contribution of $z=2$ galaxies
amounts to 10\%\ for the star forming `BX' galaxies, to  8\%\ for the rarer $K$-bright `BX' galaxies with $K_s<20$,
and to an additional 5\% for the DRGs ($J-K>2.3$).

We note that the numbers quoted in paper~II for the BX galaxies  were technically based on their stellar mass 
estimates (using a Salpeter IMF) combined with 
their metallicity, i.e. without the contribution of the ISM which was unconstrained at the time.
 However, based on \Ha\ flux measurement, \citet{ErbD_06b} estimated the gas fractions of `BX' galaxies
 to be $\sim50$\%. Thus, the contribution of the `BX' galaxies alone may be as high as 15\%. 
DRGs are generally likely to be poorer in gas.
 But, given that stellar masses $M_\star$ of high-redshift galaxies are known to be overestimated (for a Salpeter IMF) when one compares 
 $M_\star$ to their dynamical masses $M_{\rm dyn}$ \citep[see][]{ForsterSchreiberN_06a}, our estimates
  in paper~II  can be viewed as inclusive of all the baryons (gas and stars).

All in all, the total contribution of the known $z\sim2$ populations is $\sim30$\%   \citep[see also][]{PettiniM_06a}
of the metal budget corresponding to a cosmic metal density of 
\begin{equation}
\rho_{Z,\rm galaxies}\simeq 1.1\times10^6 \;\rhosun\;.\label{eq:metals:galaxies}
\end{equation}

Furthermore, since most high-redshift surveys detect galaxies to a luminosity comparable to $L^*$,
in paper~II, we estimated the contribution of the fainter galaxy population using a metallicity-luminosity
relation. We found that the $<L^*$ galaxy population could double this sum.
In other words, currently known galaxy populations at $z=2$ can only account for  30 to 60\%\ of the metals
expected, and $\ga40$\%\ are unaccounted for (see summary in Table~\ref{table:metals:galaxies}).

\subsection{Metals in absorption lines}

We separate our analysis of the metals in the IGM according to the \HI\ column density. 
The forest with $\NHI<10^{17}$~\cmsq\  is discussed in  section~\ref{section:metals:IGM},
sub-DLAs with $10^{19}<\NHI<10^{20.3}$~\cmsq\  in section~\ref{section:metals:LLS},
and  DLAs with $\NHI>10^{20.3}$~\cmsq\ in section~\ref{section:metals:DLAs}.

\subsubsection{Metals in the forest}
 \label{section:metals:IGM}
 
\begin{table*}
\caption{Metals in the forest.\label{table:metals:forest}}
\begin{tabular}{llllllllll}
\hline
Element	&	$\rho_X$			& $\Omega_X$	&	$\rho_{Z}$	& 	$\Omega_{Z}$	&	$\overline Z_b$~\tablenotemark{a}
		&  $f$	&	Method~\tablenotemark{b} & Ref~\tablenotemark{c}	& UVB	 \\
	&	(\rhosun)			&		& 	(\rhosun)	&			&		&  (\%)\\
	\hline
C	&	$3.1\times10^4$		&	$2.3\times10^{-7}$ & $3\times10^5$	&	$2.2\times10^{-6}$	&  0.0026	&  8 	&	\CIV\ HPOD &  (1) & soft\\
C	&	$4.9\times10^4$		&	$3.6\times10^{-7}$ & $4.7\times10^5$	&	$3.4\times10^{-6}$	&  0.0041	&  13 	&	\CIV\ HPOD &  (2) & hard\\
Si	&	$4.6\times 10^{4}$	&	$3.4\times 10^{-7}$& $1.2\times 10^{6}$&	$8.6\times10^{-6}$	&  0.0100	& 30 	&	\SiIV\ CPOD & (3)	& soft\\
O	&	$2.4\times 10^{5}$	&	$2.0\times 10^{-6}$& $4.6\times10^5$	&	$3.4\times10^{-6}$     &  0.0040	&  12 	&	\OVI\ 	&	(2) 	& hard \\
O	&	$7.0\times 10^{5}$	&	$5.0\times 10^{-6}$& $1.2\times10^6$	&	$8.4\times10^{-6}$    &  0.010	&  30 	&	\OVI\ 	&	(2) 	& soft \\
O	&	$1.7\times 10^{5}$	&	$2.3\times 10^{-6}$& $5.2\times10^5$	&	$3.9\times10^{-6}$	&  0.0046	&  13	 &	\OVI\	&	(4)	& hard	\\
\hline
Summary &	$3.5\times 10^{5}$	&	$2.5\times 10^{-6}$& $5.8\times10^5$	&	$4.2\times10^{-6}$	&  $<$0.0050	&  $<$15--30 	\\
\hline
\tablenotetext{a}{At $z=2$, all the baryons are in the forest,
and  the mean baryonic metallicity $\overline Z_b$ corresponds the  mean metallicity of the IGM (by mass).}
\tablenotetext{b}{HPOD: pixel optical depth using \HI; CPOD: pixel optical depth using \CIV.}
\tablenotetext{c}{(1) \citet{SchayeJ_03a}, (2) \citet{SimcoeR_04a}, (3) \citet{AguirreA_04a}, (4) \citet{BergeronJ_05a}.}
\end{tabular}
\end{table*}
 
In the low-density IGM (with $\NHI<10^{17}$~\cmsq),  C, Si, and O, as traced by \CIV, \SiIV and \OVI, are useful probes of 
 the metals contained in the IGM \citep[e.g.][]{SongailaA_01a,AguirreA_02a,SchayeJ_03a,PettiniM_03a,SimcoeR_04a,AguirreA_04a,SongailaA_05a}.  

Several approaches have used to estimate the metal content of the IGM using these ions.
For instance,  the   ratio of \CIV\ to \HI\ using the pixel optical depth method \citep[e.g.][]{AguirreA_02a}
can be converted into carbon abundances given an ionization correction \citep[e.g.][]{SchayeJ_03a}. 
These can be computed as a function of density $n_{\HI}$ for a given UV background (UVB) model using codes such as CLOUDY.
Hydrodynamical simulations can provide interpolation tables of the density (and temperature) as a function of the \lya\ optical depth (and redshift).
This was done by~\citet{SchayeJ_03a},
 under several assumptions regarding the UVB model used in generating the ionization corrections. 
  For a UVB from \citet{HaardtF_01a} including quasars and galaxies (hereafter `soft UVB'), they found
\begin{eqnarray}
\Omega_{C,\rm IGM}&\simeq&2.3\times10^{-7}\;10^{\CH+2.8}\;\left (\frac{\Omega_b}{0.045} \right ) \nn \\
\rho_{C,\rm IGM}&\simeq&3.1\times10^4\;10^{\CH+2.8} 
\end{eqnarray} 
This corresponds to a metallicity contribution of $\rho_{Z,\rm IGM}\simeq 3\times10^5\;\rhosun$, or only about 8\%\ of the metal budget 
This study had a threshold at $\tau_{\HI}\sim 1$, corresponding to $\NHI \;\sim\; 10^{13.5}$~\cmsq.
The work of \citet{SimcoeR_04a}, based on line fitting and on different assumptions, gave similar results ([C/H]=$-2.2$)
 and would be almost identical ([C/H]=$-3.1$) under similar assumptions about the UVB.
This study had a threshold at $\NHI>10^{14}$~\cmsq.

As for silicon abundances, the firmest estimates are provided by
\citet{AguirreA_04a}, who studied the forest metallicity by analyzing \SiIV\ and \CIV\ pixel optical depth
derived from high quality Keck and VLT spectra. They find that [Si/C] ranges from [Si/C] $\simeq 0.25$ (for a very soft UVB) to 
[Si/C] $\simeq 1.5$ (for the `hard' UVB).  For a fiducial `soft' UVB model, they fit a value of [Si/C]$=0.77\pm0.05$ for gas 
in over-densities of $\delta>3$ or $N_{\HI}>10^{14}$~\cmsq. This then gives a metallicity contribution of [Si/H]$=-2.03\pm0.14$, corresponding to
\begin{eqnarray}
\Omega_{Si,\rm IGM}&\simeq& 3.4\times 10^{-7}\;10^{\SiH+2.0}\;\left (\frac{\Omega_b}{0.045} \right ). \nn \\
\rho_{Si,\rm IGM}&\simeq& 4.6\times 10^{4}\;10^{\SiH+2.0} \;\rhosun\;.
\end{eqnarray}
% 4.6 seems right using the C numbers above, but I have not double-checked the z=3 hubble constant and \rho_crit.
Since about 3--5\%\ of type~II supernova metal production is Si \citep{SamlandM_98a}, this corresponds to
 \begin{eqnarray}
\rho_{Z,\rm IGM}&\simeq& 1.16\times 10^{6} \frac{y_{\rm Si}}{0.04}\;\rhosun,\label{eq:metals:IGM:Z:Aguirre}\\
\overline Z_b &\simeq& 0.010,\nn
\end{eqnarray}
i.e. $\sim$30\%\ of the metal budget. 

This result is   highly sensitive to the hardness of the assumed UVB,  since softening the UVB (for example) lowers   both the inferred [C/H] and the inferred [Si/C].
However, the UVB hardness has the opposite effect on [O/C] and [O/H], so they are also quite useful to examine.

Using a `hard' UVB, \citet{SimcoeR_04a} measured [C,O/H]   from which they
infer $\Omega_{\rm O}\simeq 1.4\times 10^{-6}\;h^{-1}\;\simeq 2.0\times 10^{-6}\;h_{70}^{-1}$
using [O/C]$\simeq 0$ set by the UVB.
The corresponding density of Oxygen is $\rho_{\rm O} \simeq 3.4\times10^5~\rhosun$, or
 \begin{eqnarray}
\rho_{Z, \rm forest}&\simeq&4.6\times10^5\frac{y_{\rm O}}{0.60}\;\rhosun,, \label{eq:metals:IGM:O:Simcoe} \\
\overline Z_b &\simeq& 0.004,\nn
\end{eqnarray}
about 12\%\ of the metal budget.  Using a `soft' UVB gives [O/C] $\simeq 0.5$ which
is more in line with the relative abundances seen in other metal-poor environments such as halo stars \citep{Cayrel_04a},
 and  increases the {\em median} [O/H] by $\sim 0.15$ dex, but a calculation of $\Omega_{\rm O}$ under this assumption was not provided. 
  If the [C/H] values of~\citet{SchayeJ_03a} are used with [O/C]=0.5 
  (which is consistent with results using the pixel optical depth technique; see Dow-Hygelund et al., in prep.), we would obtain
$\bar Z_b \simeq 0.01$, or $\sim 30\%$ of the metal budget.

Instead of looking at [O/H], \citet{BergeronJ_05a} searched
for \OVI\ where the identification of the systems is done with \CIV, i.e.  independently of $N_{\HI}$. 
These studies found two populations of \OVI\ absorbers.
The first population is 
 metal poor (with [O/H]$\simeq-2.0$) and has narrow line width ($b\la12$~\kms), indicative of photoionization.
 The second (and new) population is much more metal rich with [O/H]$\simeq-0.33$, and has larger line widths.
Globally, they found that $\Omega_{\OVI}=3.5\times 10^{-7}$, of which the metal-rich population contributes 35\%. 
Using a ionization correction $\OVI/$O=0.15 (assuming a hard UVB),   \citet{BergeronJ_05a} infer a Oxygen density $\Omega_{\rm O}=2.3\times 10^{-6}$,
which corresponds to: 
\begin{eqnarray}
\rho_{Z,\rm forest}&\simeq& 5.2\times10^5\frac{y_{\rm O}}{0.60}\;\rhosun, \label{eq:metals:IGM:O:Bergeron} \\
\overline Z_b&\simeq&0.0046,\nn
\end{eqnarray}
or 13~\%\ of the metal budget, similar to the estimate of \citet{PettiniM_06a}. 
%This
%is still short of closing the missing metals  by a factor of 3-4.

The  results from the literature are summarized in Table~\ref{table:metals:forest}.
The \lya\ forest mean metallicity (by mass) is $\overline Z_b \simeq 0.005$--0.010 (depending on the UVB model assumed and tracer element used) and
 indicates that it holds $\sim 15-30\%$ of the $z=2$ metal budget \citep[see also][]{PettiniM_06a}. 
Using carbon as a tracer leads to a somewhat smaller estimate .
%Si is overabundant for all UVB backgrounds, and yields a mean metallicity accordingly larger by a factor of $\sim2$.
%\citet{QianY_05a} have argued that the surplus Si  have been produced by pair instability SN
%of very massive stars (VMSs) at $z>15$.

Assuming that the intergalactic metal budget is dominated by warm ($\sim 10^4\,$K) photoionized gas,
the contribution of the forest is 15 (30)\%\  depending on the UVB.
If the UVB were a bit harder (or softer), it would decrease (increase) slightly 1 or 2 elements, 
but the other element(s) would increase (decrease). 
In particular, in order to have a UVB that yields a [Si/O] ratio consistent with type~II SNe, 
there is little room to change the 15-30\%\ contribution.

However, it is important to note that if a significant reservoir of metals is hidden in hot ($\ga 10^5\,$K) 
collisionally ionized gas, these would evade detection in \CIV\ and \SiIV, 
so that the ionization corrections employed in the calculations cited above would underestimate the true metal content.
 For example, in simulations including feeback by \citet{OppenheimerB_06a}, 
 heating of the gas hides a significant fraction of the carbon mass, so that the ionization fraction of carbon is than in \citet{SchayeJ_03a}
by a factor of almost three.  This might bring the carbon metallicity more 
in line with silicon and oxygen; which might be affected by prevalent hot gas to a lesser (though currently unquantified) degree.

\subsubsection{Lyman Limit Systems (LLS)}
\label{section:metals:LLS}

Lyman limit systems (LLS) with $\NHI>10^{17}$~\cmsq\ are prime
candidates for harboring the missing metals. Indeed, some of them are
highly ionized (because they have a lower \HI\ column density and might
therefore not be sufficiently self-shielded), and if a small fraction
of LLSs are metal rich   \citep[as already seen in][]{CharltonJ_03a,DingJ_03a,MasieroJ_05a}, they could contribute significantly to the metal
budget  \citep[e.g.][]{PerouxC_06a,ProchaskaJ_06a}.

 However, no model-independent constraints exist for LLS
  with $10^{17}<\NHI<10^{19}$~\cmsq, but progress is
 being made for absorbers with $10^{19}<\NHI<10^{20.3}$~\cmsq, also
 called sub-DLAs, since damping wings are clearly visible at such HI
 column densities \citep[e.g.][]{PerouxC_06a}.

Based on direct measures of the neutral gas mass and metallicity of a
sample of sub-DLAs, \citet{KulkarniV_06a} calculated the amount of
metals of these systems to be $Z_b=6.4\times10^{-4}$, or $\Omega_Z\simeq
5.1\times10^{-7}$ using no ionization correction $f$, i.e. $f=1.0$ where  $f=1/(1-x)$ for an ionization fraction $x$.
This corresponds to:
\begin{equation}
\rho_{Z, \rm sub-DLAs}\ga 6.9\times10^{4}\left(\frac{f}{1.0}\right)\frac{Z}{0.15 Z_\odot}~\rhosun ,\label{eq:metals:subDLA}
\end{equation}
or about 2\% of the metal budget.
Eq.~\ref{eq:metals:subDLA} is a lower limit since  $x=0$ was assumed.
 However,  \citet{KulkarniV_06a} 
emphasize that $x$ is model dependent and that different
sub-DLAs might have very different ionization fraction even at similar \NHI\ (see
\citet{DessaugesM_06a} for a sample of 13 ionized fraction estimates).

Indeed, \citet{ProchaskaJ_06a} deduce $x$=0.9 for one of their sub-DLA
 for which they run photo-ionisation modelling and assume $x$=0.1 for
 the other one. Based on this single measure, they estimated the
 amount of metals in  sub-DLAs (i.e. assuming $x$=0.9) to be $\Omega_Z\simeq 5\;10^{-6}$ or:
\begin{equation}
\rho_{Z, \rm sub-DLAs}\la 6.8\times10^{5}\left(\frac{f}{10}\right)\frac{Z}{0.1 Z_\odot}~\rhosun ,
\end{equation}
or 17\%\ of the metal budget.  This result used a mean metallicity for
the entire population of $0.1\;Z_\odot$ from \citet{PerouxC_03b}.

Given the potentially large reservoir of metals in sub-DLAs,
  progress in this field is advancing rapidly \citep{PerouxC_06c,ProchaskaJ_06a}.

\subsection{Damped \lya\ absorbers (DLAs)}
\label{section:metals:DLAs}

\begin{table*}
\caption{Metals in the IGM.\label{table:metals:IGM}}
\begin{tabular}{lllllll}
Class					&	 $\rho_{Z}$	& 	$\Omega_{Z}$	&	$\overline Z_b$	&  $\rho_Z/\rho_{Z,\rm tot}$~\tablenotemark{a} & Ref~\tablenotemark{b} & Note	\\
					&	($\rhosun$)	&			&	($Z_\odot$)		& (\%)	&	&\\
\hline
 \hline
DLA ($\NHI>10^{20.3}$)			&  $2.0\times10^{5}$	&  $1.5\times10^{-6}$  	& 0.0018	& 5.0	& 1 \\
sub-DLAs ($10^{19}<\NHI<10^{20.3}$)  	& $6.8\times10^{5}$	&  $5.0\times10^{-6}$   &  0.0060     & $<$17  & 2 & with $x=0.9$\\
sub-DLAs ($10^{19}<\NHI<10^{20.3}$)  	& $6.9\times10^{4}$	&  $5.1\times10^{-7}$   &  0.0006     & $>$2  & 3  & with $x=0.0$\\
 LLS ($10^{17}<\NHI<10^{19}$)  		&  ...  		&  ...   	&  ... 	& ...		&	& \\ 
Forest ($\NHI<10^{17}$)  		& $5.8\times10^5$	&  $4.2\times10^{-6}$ 	&  0.005	& $<$15 & 4 \\
\hline
IGM Metals: Total			& $1.5\times10^4$	&  $1.1\times10^{-5}$   &  0.011	&  $<37$ & excluding LLS\\
\hline
\end{tabular}
\tablenotetext{a}{References:  (1)  Eq.~\ref{eq:metals:DLA}, 
(2) \citet{ProchaskaJ_06a}, (3) \citet{KulkarniV_06a}, (4) Table~\ref{table:metals:forest}}
\end{table*}

Because (i) $\Omega_{\HI}$ appears to evolves by at most a factor of two
  to redshift zero \citep{RosenbergJ_03a,ZwaanM_05a}, and that (ii)
  the metallicity evolution in DLAs
  \citep{ProchaskaJ_00a,KulkarniV_02a,KulkarniV_05a,ProchaskaJ_03b} is
  much milder than the metallicity evolution seen in galaxies
  \citep{LillyS_03a}, one is forced to conclude that DLAs `do not
  trace everything', but are just tracing the gas in the same physical
  conditions at all redshifts.

Regardless of the origin of the gas, from an observational standpoint,
DLAs are twenty times more metal rich than the forest
\citep{PettiniM_99a,ProchaskaJ_97a,VladiloG_00a,ProchaskaJ_03b}.  But
given than they account for  2--3\%\ of the baryons
\citep{PerouxC_03a} with $\Omega_{\rm
DLA}=1.5\times10^8\;h_{70}^1\rhosun$, only 5\%\ the metals, i.e.
\begin{equation} 
\rho_{Z,\rm DLA}=2\times10^5\rhosun\;,\label{eq:metals:DLA} 
\end{equation} 
are in DLAs given their mean metallicity of $-1.15$ or 0.07$Z_\odot$
\citep[e.g.][]{PettiniM_99a,KulkarniV_05a}.~\footnote{ \citet{FoxA_07a} discussed 
 the contribution of the hot gas in DLAs probed by \OVI\ to the metal budget. 
 This amounts to $>0.7$\%\ of the metal budget and could be higher depending on
 the ionization correction.}
  
  There seem to be little amount of dust in DLAs as measured directly
from depletion pattern \citep[e.g.][]{DessaugesM_02a} or from the
attenuation of the QSO light
\citep[e.g.][]{MurphyM_04c,YorkD_06a,WildV_06a}. The latter method indicates
that $E(B-B)<0.01$ in DLAs.  However, they have been claims that dusty
DLAs would be missed from optically selected quasar samples
altogether as argued by  \citet{VladiloG_04a}. 
These authors find that the missing
fraction of DLAs is a strong function of the limiting magnitude of the
quasar sample and they concluded that (i) we might be missing as much
as 30-50\%\ of  $\Omega_{\HI}$ for QSO surveys down to
$r=20.5$, and (ii) averaged DLA metallicities could be 5 to 6 times
higher than currently observed.
Up to now, the magnitude range $r=19.5$ to $r=20.5$ has been largely unexplored, 
and it is where most of the   bias due to dust could be more significant.

Recently, \citet{Herbert-FortS_06a} identified from SDSS-DR3
a large sample of 435 metal strong QSO absorption line selected using EW(Zn),
with properties similar to the metal strong DLA of \citet{ProchaskaJ_03c}.
From their sample, they infer a global fraction of metal strong DLAs of 5\%\ down to $r=19.5$.

In order to estimate the additional contribution to $\rho_{Z}$ from metal-rich dusty DLAs, we examine the various
possibilities.
In  the scenario of \citet{VladiloG_04a},  the averaged metallicity of DLAs should be 5  times higher than the observed 
mean DLA metallicity of $Z/Z_\odot\simeq 0.07$.
Given their estimate of a 'dusty DLA' fraction missed in current surveys of 30\% (by mass),
the metallicity of this populations might be close to be around solar (but not much higher).
If we take their metallicity to be $Z\sim0.5Z_\odot$,  7 times that of  `traditional' DLAs,
 the amount of metals in this population of dusty DLAs is estimated to be:
\begin{eqnarray}
\rho_{Z,\rm DLA+}&=& 30\%\cdot\Omega_{\rm DLA}\cdot 0.5Z_{\odot} \nn\\
&\simeq& 4.7\times 10^5\;\frac{f_{DLA\rm rich}}{0.30}\frac{Z_{DLA\rm rich}}{0.5}~\rhosun\,, \label{eq:dustyDLAs}
\end{eqnarray}
 or about 12\%\ of the metal budget; a non-negligible contribution.
If the fraction of metal rich systems is closer to $\sim$5\% \citep{Herbert-FortS_06a} with a mean metallicity of $0.3\;Z_\odot$,
 then Eq.~\ref{eq:dustyDLAs}
is reduced by a factor of 10, and the contribution of dusty DLAs is not significant.
This population of dusty DLAs could contribute significantly to the metal budget.

\vspace{1cm}
In Table~\ref{table:metals:IGM}, we summarize the metal budget for the IGM.
 The sum for DLAs, sub-DLAs and the forest reaches about $<35$ to $<$40\%, which is barely what is required to close the metal budget.

\subsection{QSOs and AGN feedback}

 The contribution of QSOs to the cosmic metal budget is usually not included due to their very low
number  density, $n_{\rm QSO}\la 10^{-6}$~Mpc$^{-3}$ \citep[e.g.][]{CroomS_04a}. 
But, QSOs appear to have  
reached solar metallicity \citep[e.g.][]{Hamann_02a,HamannF_04a} based on independent analyses 
of quasar broad emission lines and intrinsic narrow absorption lines \citep[see also][]{DodoricoV_04a}.
Given that their life time is relatively
short ($\tau\la 10$~Myr, i.e. duty cycle large $\ga 100$), 
their contributions can be much larger
 if they contain large amount of gas. From CO observations,
 it has been noted that QSOs contain less gas than SMGs by a factor of $\sim 4$ \citep{GreveT_05a},
 so we expect their contribution to be less than that of the SMGs.
Typically, QSOs  have $\sim\;10^{10}$~\msun\ of gas \citep[e.g.][]{HainlineL_04a}.
We find that, the comoving amount of metals in QSOs is then:
\begin{equation}
\rho_{Z,\rm QSO}=  2\;\times\;10^{4} \frac{n_{\rm QSO}}{10^{-6}}\frac{f}{100}\frac{M_{\rm gas}}{10^{10}}\;\rhosun \,\label{eq:metals:qsos}
\end{equation}
or less than 1\%\ of the expected metals.

Thus, even if QSOs expel most of their gas, they cannot contribute to the metal budget at more than the percent level.
In fact,
\citet{NicoleN_06a} studied the kinematics of the outflow powered by the radio galaxy MRC1138$-$262, and
estimated an outflow rate 300-400~\msun~yr$^{-1}$. From the outflow rate, they estimated
 the contribution to $\rho_Z$ from AGN feedback  
assuming that all QSOs undergo a powerfull radio phase similar to that of MRC1138$-$262 (using a duty cycle of 300).
They found that they could contribute to 0.1---30$\times10^4$~\rhosun\ of metals, i.e. at most 10\%, and the most
likely value is $1\times10^4$~\rhosun\ or 0.3\%\ given that QSOs outflows are not likely to expel more metals
than their gas reservoir (Eq.~\ref{eq:metals:qsos}) .
We thus view that AGN feedback is unlikely to contribute significantly to the enrichment of the IGM,
but will significantly affect the evolution of massive objects  \citep[e.g.][]{CrotonD_05a,NicoleN_06a,BestP_06a}

\section{How many metals are ejected from small galaxies?}
\label{section:alternative}

In this section, we use simple analytical calculations based on the effective yield $y_{\rm eff}\equiv Z/\ln(1/\mu)$
(Eq.~\ref{eq:effective:yield}) to argue that, at $z=2.$, 25-50\%\ of the metals are `outside'~\footnote{By this we mean
not in stars, and not in the ISM. The metals could be in the halos of galaxies or in the IGM proper.} galaxies.
First, we test our method at $z=0$ in section~\ref{section:ejected:z0}, and then repeat it at $z=2.5$
in section~\ref{section:ejected:z2}.

%This component can be detected using absorption lines only if it is not hotter than $\la\;10^6$~K.

\subsection{The contribution of metals lost from galaxies}
\label{section:ejected:intro}

\begin{figure}
%\centerline{\includegraphics[width=80mm]{figs/yeff_plots_compare.Garnett3.eps}}
\centerline{\includegraphics[width=80mm]{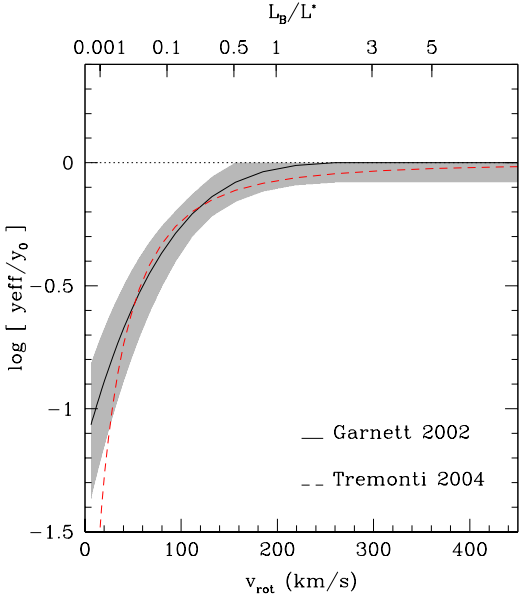}}
\caption{Effective yield ($y_{\rm eff}$) normalized to the value expected
from a close box model $y_0$. The solid line shows the relation
parameterized by \citet{GarnettD_02a}, and the dashed line shows the relation
obtained by \citet{TremontiC_04a} using SDSS data. The shaded area
represent the allowed range from the observations. }
\label{fig:effective_yield}
\end{figure}

At $z=0$,   evidence for enriched material being ejected by galaxies comes from the mass (or luminosity)--metallicity
relation \citep[e.g.][]{GarnettD_02a,PilyuginL_04a,TremontiC_04a}. Essentially any chemical model would predict that
a galaxy (as it turns its gas into star) reaches solar metallicity \citep[e.g.][]{EdmundsM_90a} even
with infall of primordial gas. Once the energy of SN is larger than the gravitational energy of the gas,
the remaining gas could be expelled, resulting in a mass--metallicity relation since the gravitational energy
will depend on the mass \citep{LarsonR_74b}.

Another signature of gas losses come from the   effective yield (Eq.~\ref{eq:effective:yield}), which
measures how far a galaxy is from a `closed-box'  evolution. A  galaxy that has  evolved
as a closed-box would obey a simple linear relationship between gas metallicity and gas fraction, the slope
being the effective yield, i.e. $Z\;=y_{\rm eff}\;\times\;\ln(1/\mu)$ (Eq.~\ref{eq:effective:yield}).

Fig.~\ref{fig:effective_yield} shows the effective yield as a function of rotational velocity $V_c$.
The solid line shows the relation
parameterized by \citet{GarnettD_02a}, and the dashed line shows the relation
obtained by \citet{TremontiC_04a} using SDSS data. The shaded area
represent the allowed range from the observations.

This effective yield has two important properties relevant for this paper.
First, as it has been shown many times \citep[e.g.][]{EdmundsM_90a,DalcantonJ_06a}, 
   outflows can lower $y_{\rm eff}$ effectively. 
Second, how much the effective yield has departed from the closed-box expectation (given by true yield)
is \citep[e.g.][]{GarnettD_02a,TremontiC_04a,PilyuginL_04a, BrooksA_06a} a 
measure of the {\it minimum} amount of metals that were 
lost in the last star-burst driven outflow, since, technically,
inflows can increase $y_{\rm eff}$ \citep{KoppenJ_99a} after an outflow episode.

It can be argued that the ratio between the effective yield  $y_{\rm eff}$ and the true yield 
$y_0$ is a measure of the lost metals
 by approximating the effective yield 
 when  the gas fraction $\mu\;\equiv\;M_{\rm gas}/(M_{\rm gas}+M_{\star})$  is close to 1,
 in which case:
\begin{eqnarray}
y_{\rm eff}&=&\frac{M_Z}{M_{\star}}\label{eq:yeff:approx}
\end{eqnarray}
where $M_Z$ is the mass of metals in the gas ($=Z\;M_{\rm gas}$). 
One sees that the effective yield is independent of how much
gas is in the system, but any process that removes metals (as in metal-rich outflows) 
will  reduce $y_{\rm eff}$ in proportions to the metals lost
since  Eq.~\ref{eq:yeff:approx}  is linearly proportional 
to the mass of metals in the gas phase.
 
One can show more directly that    the ratio $y_{\rm eff}/y_0$ is a measure of the
mass fraction of the lost metals (see Appendix~\ref{section:appendix:yeff}).
A  galaxy that experienced an outflow episode has a remaining mass of metals $M_Z^w$.
The  mass of metals lost ($M_Z^{\rm lost}$) 
with respect to that of the closed box evolution ($M_Z^{cb}$), normalized to the mass of metals left $M_Z^w$ is
(Eq.~\ref{eq:appendix:yeff}):
\begin{eqnarray}
\frac{M_Z^{\rm lost}}{M_Z^{\rm w}} &=&\frac{M_Z^{\rm cb}-M_Z^{\rm w}}{M_Z^{\rm w}} \nn\\
&=&\frac{y_0\;  M_{\rm bar}^{\rm cb}}{y_{\rm eff}\;  M_{\rm bar}^{\rm w}}-1 ,
\end{eqnarray}
where $M_{\rm bar}^{\rm cb}$ and $M_{\rm bar}^{\rm w}$ are the baryonic masses 
for the closed box evolution and for the wind scenario, respectively.

In general, the baryonic mass is not affect by outflows, i.e. $  M_{\rm bar}^{\rm cb}/M_{\rm bar}^{\rm w}\simeq1$,  and 
\begin{eqnarray}
\frac{M_Z^{\rm lost}}{M_Z^{\rm w}} 
&=&\frac{y_0}{y_{\rm eff}}-1 . \label{eq:yield:lost}  
\end{eqnarray}

From Eq.~\ref{eq:yield:lost}, one sees that
the ratio between $y_0$ 
and $y_{\rm eff}$ gives the   fraction of metals
that were lost from the system.  For instance, if
$ y_{\rm eff}/y_0 = 1/2$ then the amount of metals lost is equal to those remaining, i.e. 50\%\
of the metals produced were lost.

This property of $y_{\rm eff}$ can be used to compute the mass of metals   ejected from galaxies of a given luminosity $L_B$
\citep[e.g.][]{GarnettD_02a} as follows:
\begin{equation}
M_{\rm O, \rm lost}=12\frac{\rm O}{\rm H}(L_B)\; L_B \frac{M_\star}{L_B}\left [\frac{y_0}{y_{\rm eff}}-1\right
]\msun \;, \label{eq:Garnett:Mlos}
\end{equation}
where $\rm O/\rm H(L_B)$ is the luminosity-metallicity (\LZ) relation, 
$M_\star/L_B$ the stellar mass ratio, the factor 12 converts the oxygen ratio to a mass ratio taking into account 
the He fraction, and $y_{\rm eff}$ the effective yield. 

To illustrate Eq.~\ref{eq:Garnett:Mlos}, we compare its prediction to the observations of
 NGC1569 \citep{MartinC_02a}, where the amount of Oxygen in the outflow was measured.
This nearby dwarf galaxy has a baryonic mass of $\sim 2\times10^8$~\msun\ \citep{MartinC_98a},
a metallicity of 0.2$Z_{\odot}$, 
and entered a starburst phase 10-20 Myr ago. The cumulative effect of $\sim$30,000 supernovae
in the central region is driving a bipolar outflow seen extending to either side of the disk
in \Ha\ emission \citep{MartinC_98a} and in X-rays. Using {\it Chandra}, \citet{MartinC_02a}
measured the amount of metals in the outflow from the X-ray emitting gas, 
and found that the hot wind carries $3.5\times10^6$~\msun\ of gas which
includes $3$--$5\times10^4$~\msun\ of Oxygen.
The gas   and the current stellar masses are both uncertain in this galaxy. 
For a  gas fraction of $\sim$40, 50, \& 60\%\ and given its baryonic mass,   the effective yield is about 
0.4, 0.5, and 0.7, respectively. The corresponding amount of Oxygen in the ISM
is  $1.5$--$3\times 10^5$~\msun.
Using  Eq.~\ref{eq:yield:lost}, 
the mass of Oxygen lost   is about 4, 2.5, and 1.2 $\times10^5$~\msun.
This is larger than the direct measurement of \citet{MartinC_02a}. However,
   the   yield   reflects the cumulative effect of
all the past bursts. In fact,   \citet{AngerettiK_05a} showed that this galaxy
likely experienced three separate starburst phases, giving a mass loss
per starburst phase of $1.3$ to 0.4 $\times 10^5$~\msun. This is not far
from the    amount of metals in the current outflow found by  \citet{MartinC_02a}
given that some additional amount of oxygen can be in a much hotter ($T>>10^{6}$~K) gas
difficult to detect.

Eq.~\ref{eq:Garnett:Mlos}  can be used in combination with a luminosity function (LF)
to compute  the global amount of oxygen (per unit volume per unit magnitude)
 ejected into the IGM as a function of luminosity :
 \begin{eqnarray}
%\frac{\mathrm d M_{\rm O}}{\mathrm d V\mathrm d L} \;=\; M_{\rm O,\rm lost}(L_B) \phi(L_B). \nn\\
\frac{\mathrm d M_{\rm O}}{\mathrm d V\mathrm d {\rm mag}} \;=\; M_{\rm O,\rm lost}(M_B) \phi(M_B)\;, \label{eq:masslost:oxygen}
 \end{eqnarray}
whose  integral    with respect to magnitude gives the total 
co-moving density of oxygen lost $\rho_{\rm O}$ from the  ISM of galaxies.

In section \ref{section:ejected:z0}, we perform the integral (whoe input details are presented in the Appendix~\ref{section:appendix:yeff})   at $z=0$ using
either the $L_B$ luminosity function or the stellar mass $M_\star$ function. 
We find that both approaches give very similar results.
Furthermore, we will show that the amount of metals lost from galaxies
corresponds to what is observed in the intra-group and cluster medium
and conclude that our methodology is sound.

\subsection{Metal loss at $z=0$}
\label{section:ejected:z0}

\begin{figure*}
\centerline{
\includegraphics[width=60mm]{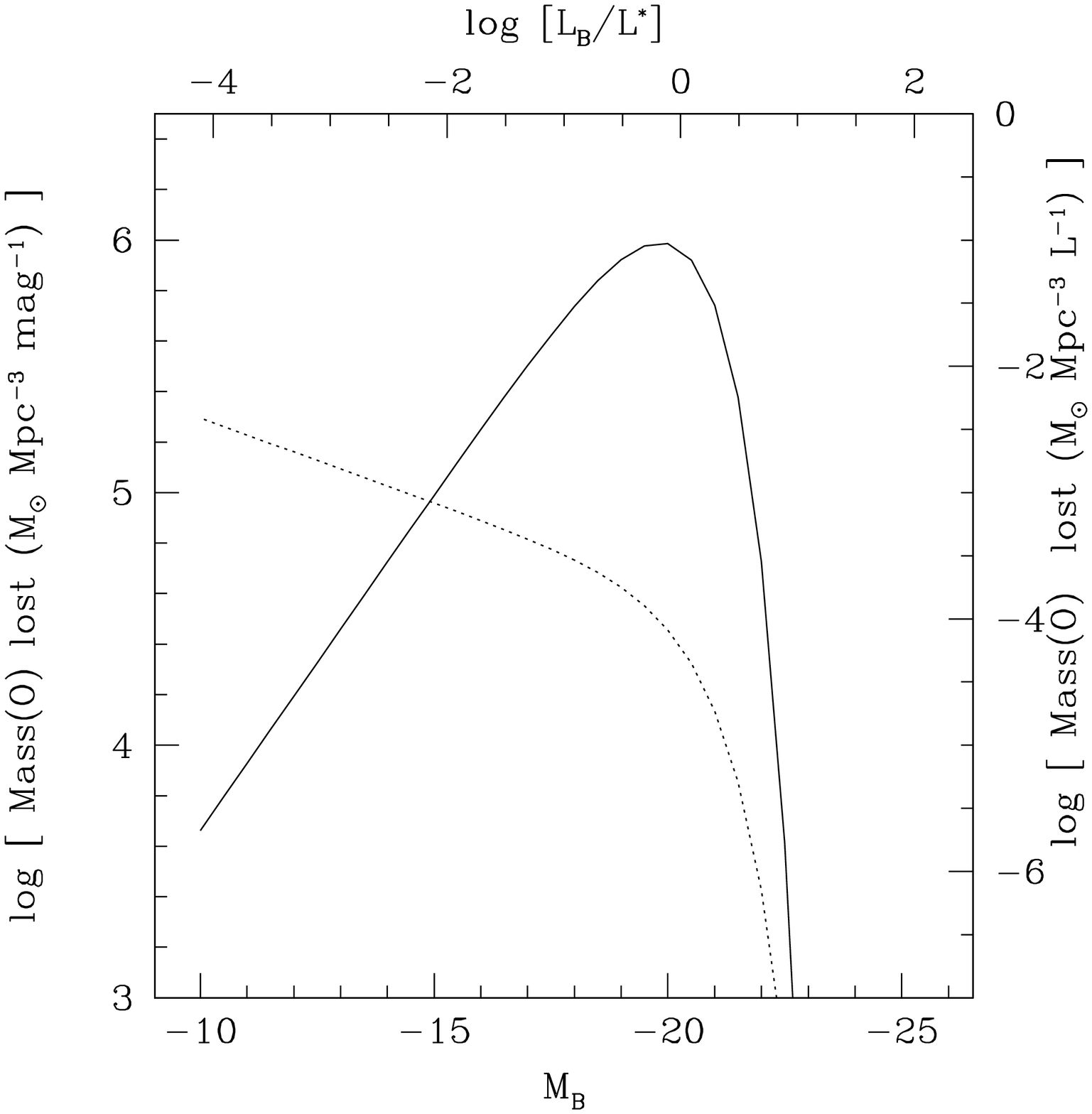}
\includegraphics[width=60mm]{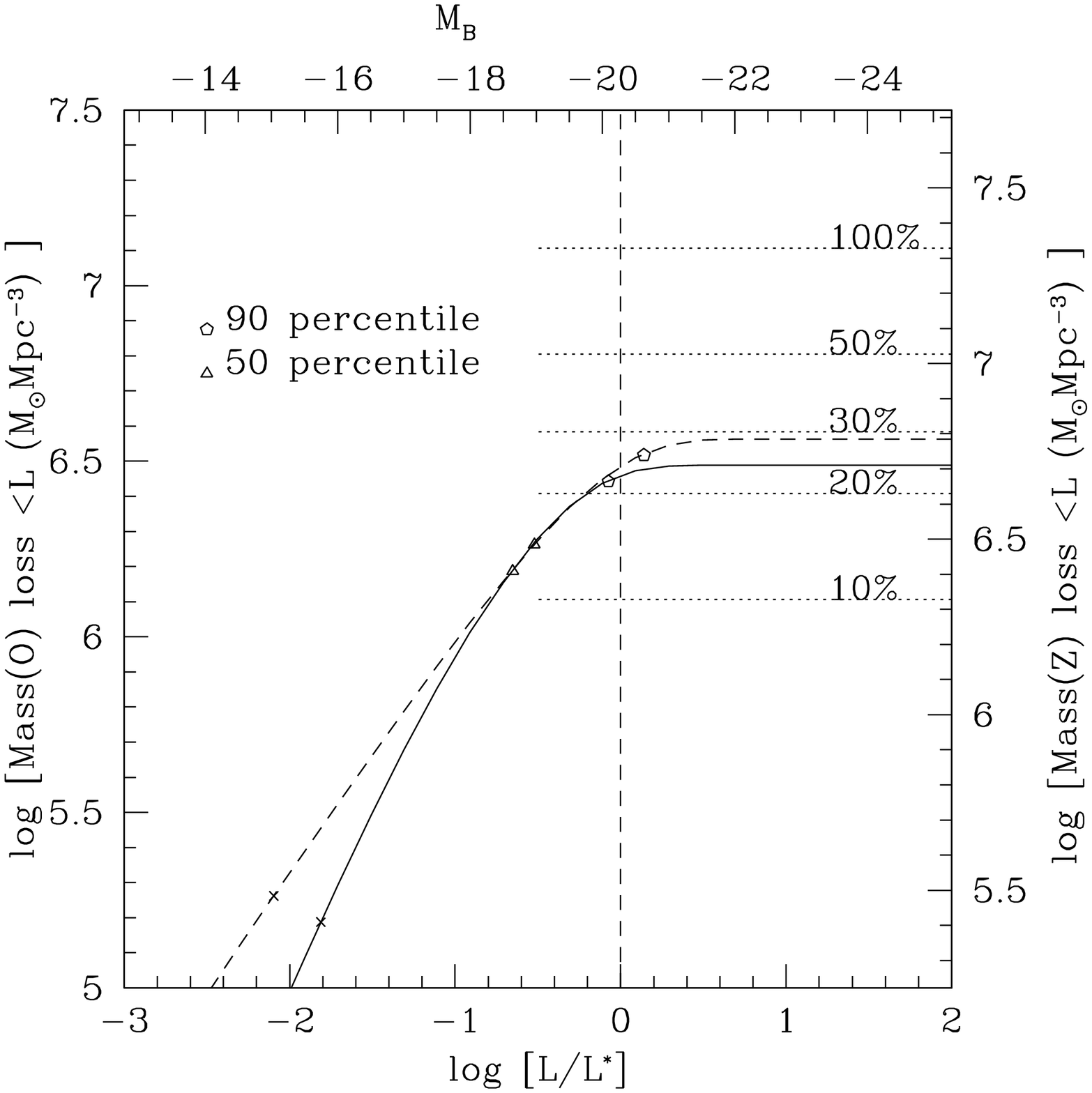}
}
\caption{{\bf Left}: the solid line (and the left $y$ scale) show 
the  amount of metals lost  
 as a function of $M_B$ using our $z=0$ effective yield calculation and Eq.~\ref{eq:masslost:oxygen}.
The dashed line (and the right $y$ scale) show the relative amount of metals lost
per unit luminosity. The solid line shows that the metals are  most likely coming out $\sim L^*_B$ galaxies.
{\bf Right}:  
The solid curve shows the cumulative integral of the solid curve shown in the left panel.
  The dashed curve show the result
if one uses $\frac{\rm O}{\rm H}$ and $y_{\rm eff}$ parameterized by \citet{TremontiC_04a}.
The 50th and 90th percentiles are shown with an open triangle and pentagon, respectively.
The 50th percentile is at $\sim 0.1L^*_B$ galaxies and shows that dwarf galaxies eject more metals into the IGM than $\ga\;L^*$ galaxies.
Percentages of the $z=0$ metal budget (Eq.~\ref{eq:massmetals:z=0}) are shown by the horizontal dotted lines.
This plot shows that the metals ejected from galaxies account  for 20--30\%\ of the metal
budget from our effective yield calculations. 
Since this amount is very close to current estimates of the observed metal mass in plasmas (see text),
we conclude that the $y_{\rm eff}$ methodology is sound.  }
\label{fig:garnett:z0}
\end{figure*}

 The solid line in  Fig.~\ref{fig:garnett:z0} (left) shows the amount of oxygen lost (i.e. ejected 
in their halos, and/or in the IGM) in
units of \rhosun\  produced by $z=0$ galaxies computed using 
Eq.~\ref{eq:masslost:oxygen}.
This solid curve peaks at $\sim 0.2L^*$ and
shows that the   bulk of the metals (given by the median or the peak) is ejected by sub-L$^*$ galaxies.
The dotted line show the  same but
 per unit luminosity (right axis) instead of magnitude,
  for comparison with \citet{GarnettD_02a}. 

The solid line in  Fig.~\ref{fig:garnett:z0} (right) shows the cumulative distribution
of $\rho_{\rm O, \rm lost}$  using \citet{GarnettD_02a}'s parameterization (Eq.~\ref{eq:Garnett:yeff}) of $y_{\rm eff}$.
 We also show  the 50th (90th) percentile as open triangle (pentagon).
The dashed curve shows the result of the integration using $y_{\rm eff}$ parameterized by \citet{TremontiC_04a} (Eq.~\ref{eq:Tremonti:yeff}).
From this plot, one sees that the total amount of oxygen ejected from galaxies is
\begin{equation}
\rho_{\rm O, \rm ejected}(z=0) \approx \rhoOlocal\;. \label{eq:Mlos:O:z0}
\end{equation} 
% metals_cumulativ_plot 'Garnett' 'Tully00' 'Garnett' 6 'z02dF'

 The oxygen yield is about 60\%\ of all metals \citep{SamlandM_98a}, so
 we find that the total 
co-moving density of metals lost from the  ISM of galaxies:
\begin{equation}
\rho_{Z, \rm ejected}(z=0) \approx \rhoZlocal \label{eq:Mlos:z0}.
\end{equation}
This correspond to $\sim$25\%\ of the $z=0$ metal budget (Eq.\ref{eq:massmetals:z=0})
or to $\overline Z_b=0.045$.
  The  ingredients that went into Eq.~\ref{eq:Mlos:z0} are  presented in Appendix~\ref{appendix:z0}.

One can perform the same calculations directly in terms of stellar masses $M_\star$ 
using stellar mass functions.
Using the results of \citet{TremontiC_04a} for $y_{\rm eff}$ (Eq.~\ref{eq:Tremonti:yeff}) and
for  the mass-metallicity ($M_\star$--$Z$) relation (Eq.~\ref{eq:Tremonti:LZ}),
 the baryonic Tully-Fischer relation from \citep{BellE_01a},
and  the stellar mass function from  \citet{ReadJ_05a},
we find that the comoving oxygen density is $\rho_{\rm O, \rm ejected}(z=0) \simeq 2.0\times 10^6\;\rhosun\;$, 
corresponding to a metal density of
%metals_cumulativ_plot 'Tremonti' 'Mbar' 'Tremonti' 1 'z0Read'
\begin{eqnarray}
\rho_{Z, \rm ejected}(z=0) &\simeq &3.3\times 10^6\;\rhosun\;. \label{eq:Mlos:z0:Mstar}
\end{eqnarray}
This is close to  our estimate based on the LF (Eq.~\ref{eq:Mlos:z0})
and very different assumptions.
The ingredients that went into this estimate are presented in Appendix~\ref{appendix:stellarmass}

Direct measurements of the plasma associated with galaxies (outside rich clusters)
 are still quite uncertain, but their contribution to the baryon budget
appear to be significant.
 According to  \citet{FukugitaM_98a}, the mass density of the group plasma
 is low $\Omega_{\rm gr}=0.005$ for $h=0.70$.
The metallicity of this plasma can only be measured in the brightest X-ray groups
and ranges from 0.1 solar to 0.6 solar.
According to the review of \citet{MulchaeyJ_00a},
this intra-group medium  has a mean metallicity similar to that of clusters, i.e. $1/3$ solar \citep[see also][]{FinoguenovA_06a}.
As a consequence, the amount of metals observed in groups is, 
$\rho_{Z, \rm groups}(z=0)\simeq 4.3\times 10^6~\rhosun,$
around 18\%\ of the $z=0$ metals produced by type~II supernovae.

The contribution of clusters is about half that of groups since 
the cluster mass density $\Omega_{\rm cl}$ is about half that of groups \citep{FukugitaM_98a} and both
groups and cluster plasmas have similar metallicities.

All in all, the amount of metals outside galaxies observed in the intra-group and intra-cluster medium is
\begin{equation}
\rho_{Z, \rm plasma}(z=0)\simeq 6\;\times\; 10^6~\rhosun, \label{eq:metals:plasma}
\end{equation}
around 25--30\%\ of the $z=0$ metals, close to our estimate based
on the effective yield   (Eq.~\ref{eq:Mlos:z0}).
This leads us to conclude that our procedure
 to estimate the amount of metals outside galaxies (i.e. not in stars and not
in the ISM) is sensible, given the uncertainties.
The agreement is in fact remarkable, since   the normalization of Eq.~\ref{eq:masslost:oxygen}
depends strongly on the normalization of the LF, and on the effective yield method.

\subsection{Metal lost at $z=2.5$}
\label{section:ejected:z2}

\begin{figure*}
\centerline{
\includegraphics[width=60mm]{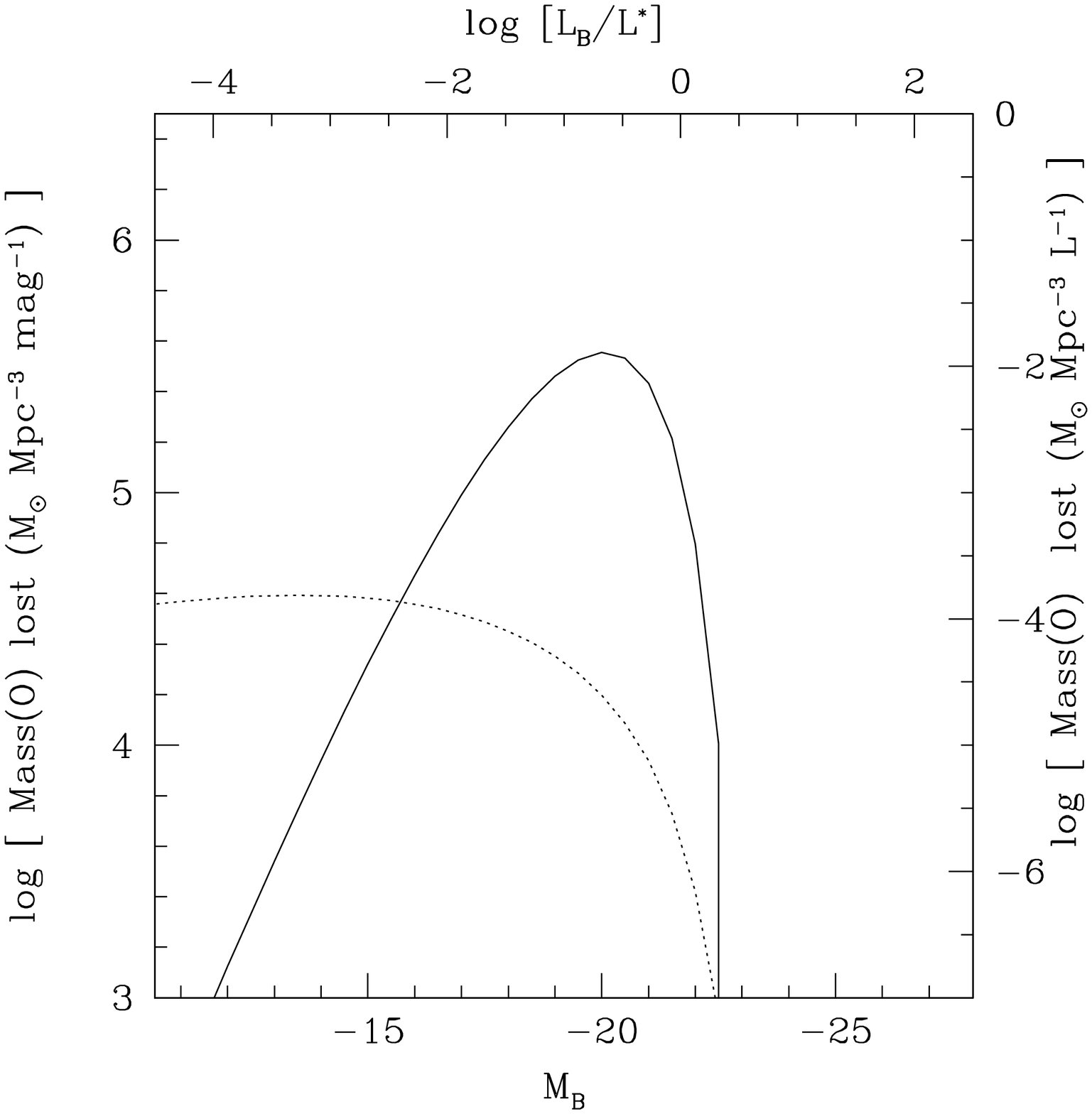}
\includegraphics[width=60mm]{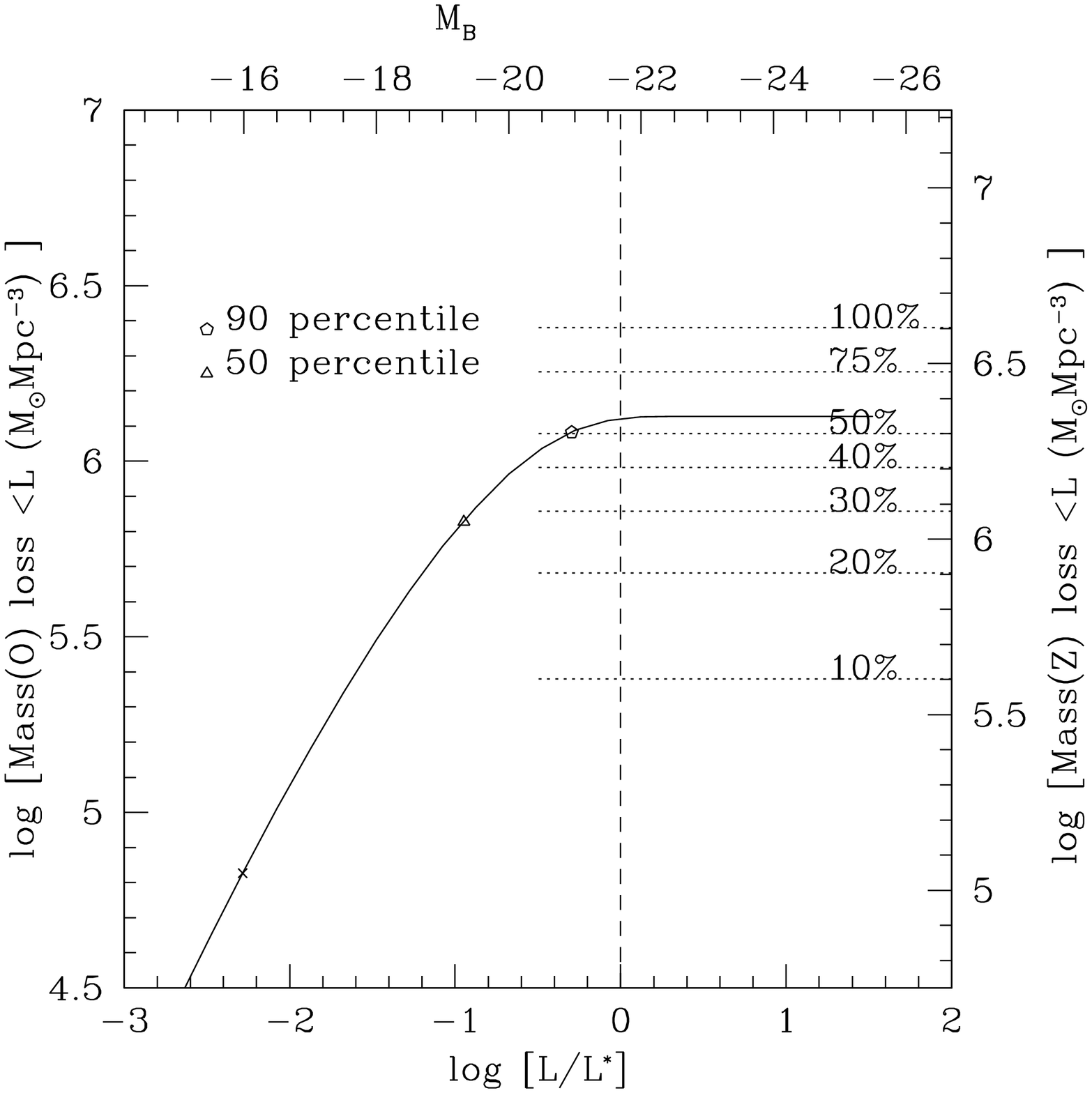}
}
\caption{Same as Fig.~\ref{fig:garnett:z0} but at $z\simeq 2.5$.
{\bf Left}: the solid line (and the left $y$ scale) show 
the relative amount of metals lost (\rhosun) 
 as a function of luminosity using our $z=2.5$ effective yield calculation (see text) and Eq.~\ref{eq:masslost:oxygen}.
{\bf Right}:  The solid curve shows the cumulative integral of the dotted curve shown in the left panel.
  The dashed curve show the result
if one uses $\frac{\rm O}{\rm H}$ and $y_{\rm eff}$ parameterized by \citet{TremontiC_04a}.
The 50th and 90th percentiles are shown with an open triangle and pentagon, respectively.
The 50th percentile is at $\sim 0.1L^*_B$ galaxies and shows that dwarf galaxies eject more metals into the IGM than $\ga\;L^*$ galaxies.
Percentages of the $z=2.5$ metal budget (Eq.~\ref{eq:massmetals}) are shown by the horizontal dotted lines.
This plot shows that about 50\%\ of the metals have been ejected from galaxies, 90\%\ of which by $\la\;L^*$ galaxies,
which is enough to close the metal budget: $>40$\%\ are missing when one accounts for $z=2.5$ galaxies (Table~\ref{table:metals:galaxies}).
}
\label{fig:garnett:z3}
\end{figure*}

We now turn towards the redshift of interest $z\simeq2.5$, and
perform a similar calculation as in section~\ref{section:ejected:z0}.
The details of
our assumptions are explained in section~\ref{section:appendix:z2}.
Briefly, using the   $y_{\rm eff}$ relation from \citet{GarnettD_02a} (Eq.~\ref{eq:Garnett:yeff}),
the \LZ\ relation  from \citet{ErbD_06a} (0.3~dex offset compared to $z=0$),
the  $B$-band luminosity function from  \citet{SawickiM_06a},
 the $B$-band Tully-Fischer relation (TFR) offset by 1~mag (see section \ref{section:appendix:z2})~\footnote{no-evolution of the $B$-band TFR 
decreases our results by a factor of $\approx 2$.},
and  a $M/L_B$ ratio equals to 2,
%metals_cumulativ_plot 'Garnett' 'highzTF' 'Erb06' 2 'z3G'
we find that
\begin{eqnarray}
\rho_{O, \rm ejected}(z=2.2) &\simeq &1.3\times 10^6\;\rhosun\;, \label{eq:Mlos:O:z2}\\
\rho_{Z, \rm ejected}(z=2.2) &\simeq& 2.1\times 10^6\;\rhosun\;, \label{eq:Mlos:z2}
\end{eqnarray}
which corresponds about $\sim 50$\%\ of the expected metals (Eq.~\ref{eq:massmetals})
and to a mean metallicity (if we spread these metals over all the baryons):
\begin{eqnarray} 
\overline Z_b &=&0.018.
\end{eqnarray}

 The solid line in Fig.~\ref{fig:garnett:z3} (left) shows the amount of oxygen lost in units of \rhosun\ produced at $z=2.5$ 
computed similarly as in Fig.~\ref{fig:garnett:z0}. 
The solid curve peaks at $\sim L^*$ and
shows that the   bulk of the metals (given by the median or the peak) is ejected by sub-L$^*$ galaxies.
The dotted line show the  same but
 per unit luminosity (right axis) instead of magnitude,
  for comparison with \citet{GarnettD_02a}. 

The solid line in  Fig.~\ref{fig:garnett:z0}(right) shows the cumulative distribution
of $\rho_{\rm O, \rm lost}$.
The 50th (90th) percentile are shown as open triangle (pentagon).
The dashed curve shows the result of the integration using $y_{\rm eff}$ parameterized by \citet{TremontiC_04a} (Eq.~\ref{eq:Tremonti:yeff}).

These plots indicate that sub-L$^*$ galaxies  have ejected enough metals ($50$\%) to close the metal budget, some (1/3) of which
is already detected in the forest according to our results in section~\ref{section:metals:IGM}, and the remainder likely being
in a hotter phase.
In particular, low mass ($L_B<\frac{1}{3}L^*(z=2)$) galaxies are responsible for 90\%\  the production of these
hot metals,  the median being at $L\sim \frac{1}{10}L^*_B(z=2)$.
Given that, at $z=0$, the mass outflow
rate is usually comparable to the SFR \citep[][for a recent review]{LehnertM_96b,HeckmanT_00a,VeilleuxS_05a}, 
and that direct evidence for winds for LBGs are numerous  \citep[e.g][]{PettiniM_01a,AdelbergerK_03a,ShapleyA_03a},
our  result may not be very surprising.

Our conclusion that   $\sim$50\%\  of the metals have been expelled from small galaxies
is very consistent with
 \citet{SimcoeR_04a} who treated the universe as the `ultimate closed box model',
and found that the fraction of the metals formed in a starburst that are ejected
from the galaxy into the IGM, $f_{\rm ej}$ is $\ga15$\%. They argue that the value
of $f_{\rm ej}$ is probably higher since gas to stellar ratio are much higher  at earlier times
than at  $z=0$. This limit $\ga15$\%\ is consistent with our estimate of $\sim50$\%.

\section{Insights from simulations}
\label{section:metals:IGM:simulations}

Several approaches have been taken to predict the cosmic metal abundances and its evolution analytically.
Early chemical models trace only the cosmic mean metallicity Z \citep[e.g.][]{PeiY_95a,EdmundsM_97a,PeiY_99a}.
 Recent chemical models \citep[e.g.][]{CaluraF_04a,CaluraF_06a} have been able to keep track of each elements
 over cosmic times, but in  typically only 3 types of galaxies, spheroids, spirals and dwarfs.
In the study of \citet{CaluraF_06a}, where they modeled spheroids and dwarfs only,
spheroid formation at $z\sim5$  produces most the IGM  metals,
such that the IGM metallicity is fairly flat from $z=4.5$ to the present.

Many numerical models have been developed in order to simulate the distribution
of the metals in the IGM. But, even in models that do include galactic winds \citep[e.g.][]{CenR_99a,AguirreA_01a,TheunsT_02a,BertoneS_05a,CenR_05a},
the predicted IGM metallicity is too high
compared to the mean metallicity of the forest, which is very low (0.005~$Z_\odot$).
For instance, \citet{BertoneS_05a} modeled galactic winds analytically in N-body simulations
and found that winds should have enriched the IGM to a metallicity of $10^{-2}$--$10^{-1.2}$,
about   10 times larger than what is observed, a conclusion reached by \citet{CenR_05a}
using very different simulations. 
Recently, \citet{BertoneS_07a} investigated the fraction of metals lost from galaxies
in galaxy formation models. They specifically looked the amount of metals that are permanently lost
from the host galaxy, and computed the distribution of metals lost as a function of virial mass $M_{\rm vir}$.
They found that at $z=0$, the 50th precentile of this distribution is at $\sim10^{11.5}$~\msun, or about
0.8dex smaller than $L^*$ (using $M_{\rm vir}^*=10^{12.3}$~\msun). This is rather close to our estimate
shown in Fig.~\ref{fig:garnett:z0} (right) of $\log [L/L^*]=0.6$. 
A preliminary comparison  between their $z=2$ result  expressed as a function of luminosity (Bertone, S. private
communication) and Fig.~\ref{fig:garnett:z3} yields very encouraging qualitative agreement.

 Keeping track of individual elements in cosmological simulations is more difficult.
 Recently, chemical models of individual elements have been attached to SPH simulations \citep{SamlandM_03a,KobayashiC_04a},
 but are often limited to  single halos.
\citet{KobayashiC_06a} combined chemical evolution models in a full cosmological tree-SPH simulations.
They find that 20\%\ of all baryons are ejected at least once from galaxies into the IGM.
Galactic winds are found particularly efficient in low mass galaxies. They argue that the origin
of the mass-metallicity relation is from galactic winds.

Using hydrodynamical simulations that incorporate metal-enriched kinetic feedback,
 \citet{DaveR_06a} concluded that  50\%\ of the $z=2$   metals are not in galaxies.
These simulations use scaling relations that arise in momentum-driven winds \citep{MurrayN_05a}
and ionization calculations to calculate the ionization fractions of atomic species given the gas density, temperature
and the ionization field.
In order to reproduce the abundances of \CIV\ from $z=6$ to $z=2$ \citep{OppenheimerB_06a} generally requires high mass loading factors
but low velocity winds from early galaxies, in order to eject a substantial metal mass into the IGM
without overheating it. In other words, momentum-driven outflows provide the best agreement with observations
of \CIV.
The  simulations of \citet{DaveR_06a} are in good agreement with several of our results. For instance, they find that 20--30\%\
of the metals are locked in stars at $z=2$, 15--20\%\ are in the shocked IGM (akin to \OVI), and 15\%\ reside is cold star-forming
gas in galaxies (akin to DLAs and sub-DLAs) and 30\%\ are in the diffuse IGM (akin to the forest). 
Essentially, they find that 40\%\ of the metals are in galaxies, in agreement with our constraints of $\ga30$ and $\la30$\%.
Half of the remaining 60\%\ is predicted to be in the diffuse IGM (although with substantial fraction in hotter gas),
and the rest are divided between the shocked IGM and the hot halo gas.

Other predictions of the metal enrichment  include 
\citet{MadauP_01a,DaveR_01a,ScannapiecoE_02a,FurlanettoS_03a,DeLuciaG_04a,ScannapiecoE_05a,PorcianiC_05a,FerraraA_05a}.

\section{Summary \& Conclusions}
\label{section:discussion}

In terms of the  $z\sim2.5$ metal budget, the dominant contributors are the galaxies, the forest and sub-DLAs:
\begin{enumerate}
\item $\ga30$\%\ and $\la60$\% of the metals are currently observed in $z\sim2.5$ galaxies (Eq.~\ref{eq:metals:galaxies}),
 according to our calculations in paper~I, and paper~II extended here and
summarized in Table~\ref{table:metals:galaxies};
\item 15--30\%\ of the metals are in the low column density IGM  with  $\log \NHI<17$,
and $\la17$\%\ are in sub-DLAs with $19<\log[\NHI]<20.3$
as estimated, in section~\ref{section:metalbudget}, and summarized in Table~\ref{table:metals:forest};
\item from our effective yield calculations (Fig.~\ref{fig:garnett:z3}), $\sim$50\%\ of the   metals are predicted to be outside galaxies.
This method was successfully tested at $z=0$, where the census of metals outside galaxies is more robust.
Furthermore, we find that low mass ($L_B<\frac{1}{3}L^*(z=2)$) galaxies are responsible for 90\%\ of these metals.
\item this last result agrees well with observations since 
 we can account for $\sim 30$--45\%\ of the metals outside the ISM of galaxies by adding the contribution
 of the forest and of sub-DLAs.
We note, that the contribution of absorbers with $17<\log[\NHI]<19$ is currently lacking, and that
this class (plus the sub-DLAs)  are potentially a good place to look
for the remaining metals as current constraints are very limited;
\item comparing the amount of metals outside galaxies at $z=0$ (Eq.~\ref{eq:Mlos:z0})  and at $z=2.5$ (Eq.~\ref{eq:Mlos:z2}),
one sees that it  has evolved by a factor of 2.
\end{enumerate}

Taken at face-value,  one could conclude that the metal budget is almost closed
since 30--$60$\%\ of the metals are in galaxies and $\la30$--45\%\ are in the IGM.
However, given the potential overlap between the various galaxy populations, we think 
 it is very unlikely that all the metals have been accounted for. 
Conservatively,  only about $>$65\%\ of the metals have been detected directly, 
equally spread between galaxies and the IGM. 

Our effective yield calculations indicate that  metal rich  outflows from galaxies  
are likely to   be the reservoir of the remaining missing metals. A significant fraction (2/3) of this is already seen in absorption lines studies
and the remaining metals are very likely in a hot phase (hotter than photoionization temperatures), a
conclusion reached by \citet{FerraraA_05a} and \citet{DaveR_06a}.

\section{Discussions} 

We end this paper by discussing  three related questions.

$\bullet$ Are the metals seen in the IGM at $z=2$ produced by $z=2$ galaxies?
 In other words, is the comoving number density of `BX' galaxies 
 large enough and is their star formation phase  lasting long enough for producing 50\%\
of the $z=2.5$ cosmic metal density?
Given the co-moving density of $z=2$ galaxies is $n\simeq2\times10^{-3}\;h_{70}^3$~Mpc$^{-3}$ \citep{AdelbergerK_04a},   their typical
SFR are $\simeq 35$~\mpy\ \citep{ShapleyA_03a},   ages of $2\times10^8$~yr \citep{ShapleyA_05a}, and metallicities of 0.5~$Z_\odot$ \citep{ErbD_06a},
 the co-moving metal density in winds from star-forming galaxies is:
\begin{eqnarray}
\rho_{Z}&=& 5.3\times10^5\left(\frac{r}{4}\right)\left(\frac{SFR}{35\hbox{\mpy}}\right)\left(\frac{\tau}{2\;10^{8}~{\rm yr}}\right)\nn\\
&& \left(\frac{Z}{0.5 Z_\odot}\right)
\left(\frac{n}{2\;10^{-3}\hbox{Mpc$^{-3}$}}\right)~\rhosun\,,\label{eq:metals:outflow}
\label{eq:metals:LBG:winds}
\end{eqnarray}
where we assumed a wind mass outflow rate $r$ of 4 times the SFR \citep{ErbD_06a}.
Eq.~\ref{eq:metals:outflow} amounts to about $\sim$10\%\ of the metal budget, i.e. much smaller than the $>40$\%\ missing, or
the 50\%\ expected to be outside galaxies~\footnote{We note that, technically, the outflow `rate' $r$ inferred
by  \citet{ErbD_06a} gives a true outflow rate of  $(1-R)\;r$ times the SFR, where $R$ is the mass fraction in massive stars ($\sim0.3$).}.
%For a normal solar true yield, $r$ is about 1--2, closer
%to the local outflow rates \citep{LehnertM_96b,HeckmanT_00a}, bringing Eq.~\ref{eq:metals:outflow} down by a factor of 2.
This is telling us  that 3--5 bursts of star formation are required per galaxy.

$\bullet$ A possibility for the remaining missing metals is that they are in a phase hotter than photoionization temperatures, at $T>5\;\times\;10^5$~K.
A natural question is then: are there enough energy in the outflows to keep some of the gas invisible?
The energy for one hydrogen atom (or baryon) in the outflow of velocity $V_{\rm outflow}$ is $\frac{1}{2}m_H\;V_{\rm outflow}^2$. 
The characterisitic outflow speed is 400--800~\kms\ \citep[e.g.][]{HeckmanT_00a,MartinC_05a} at $z=0$ and $\sim600$~\kms\ at $z=2.5$
from the shift of the the low-ionization lines to the \lya\ emission lines \citep{ShapleyA_03a}. 
Thus, the outflow carries $\sim\;2\;\left (\frac{V_{\rm outflow}}{600\;\kms}\right )^2$~keV/baryon of kinetic energy. 
Whereas, the energy necessary to heat the gas to $\sim10^6$~K is $\frac{3}{2}\;kT$, which 
is $\sim\;120\;\left (\frac{T}{10^6\;\rm{K}}\right )$~eV per baryon.
Thus, only 6\%\ (ranging from 3 to 15\% for $V_{\rm outflow}=$400--800\kms) of the kinetic energy available is necessary to heat the gas to $T\;\sim\;10^6$~K.

$\bullet$  Are the metals seen in the IGM at $z=2$ going to fall back or leave the virial radius for ever?
If galaxies evolve   with only outflows at an outflow rate always equal to the star formation rate, then 
there would always be as many metals in stars as in the IGM at all times, including at $z=0$.
Thus,  the ratio between the metal density of the IGM to the metal density in stars would have to stay constant.
But, at $z=0$, about 80--90\%\ of all the metals produced by type~II and type~Ia supernovae are locked in stars, i.e.
$<10$\%\ of the total metal density  are in plasmas outside galaxies, i.e.  that ratio is 1:9.
From this work, the metals outside galaxies amount to $\rho_{Z,\rm ejected}(z=2)\sim 2\times10^6$~\rhosun\ and 
the stellar metal density is $\rho_{Z, \rm stars}(z=2)\sim1$--$2\times10^6$~\rhosun, i.e. that ratio is 2:1 or 1:1.
Since this ratio has evolved from 2:1 at $z=2$ to 1:9 at $z=0$, we view this 
as evidence that a significant fraction of the IGM metals have cooled and fallen back into galaxies.
This  conclusion was reached using $z=0$ SDSS data by \citet{KauffmannG_06a} on very different ground,
and is consistent with the models of \citet{DaveR_06a}.
This dramatic change in proportions may be related to a shift in the dominant phase of star formation:
where, at $z=2.5$,  bursty (i.e. producing outflows) star formation dominates due to rapid accretion and/or merger,
while   more quiescent mode of star formation dominates  at smaller redshifts due to more slower accretion.
This is very reminiscent of the redshift dependence of the two   modes of gas accretion discussed in \citet{KeresD_06a}.
 At high redshifts, cold accretion along   filaments dominates, while at low-$z$, 
 the hot accreation dominates. Both modes qualitatively reproduce the rapid vs. quiescent mode of star formation.

\section*{Acknowledgments}
N.B. wishes to thank   l'Institut d'Astrophysique de Paris for its hospitality
during parts of the writing of this paper and for 
a grant  from the GDRE EARA, European Association for Research in Astronomy.
J. Brinchmann, R. Dav\'e, S. Charlot, B. Oppenheimer, and K. Finlator are thanked for discussions. 
 AA  gratefully acknowledges support from NSF Grant AST-0507117.
We thank the anonymous referee for a positive report that improved the manuscript.

%\bibliography{references}
%\bibliographystyle{mn2e}

\appendix

\section{Effective yield calculations}
\label{section:appendix:yeff}

In Appendix~\ref{appendix:yeff:fraction}, we justify Eq.~\ref{eq:yield:lost}
which is central to this paper 
and describe the behavior of 
Eqs.~\ref{eq:Garnett:Mlos} and  \ref{eq:masslost:oxygen}
in Appendix~\ref{appendix:dependence}.

\subsection{Fraction of metals lost}
\label{appendix:yeff:fraction}

One can show that the fraction of metals lost (with respect to a closed box
model) is proportional to $y_0/y_{\rm eff}-1$.
Consider a galaxy that 
has evolved for a given amount of time, untill some gas fraction $\mu$ remains,
 and experienced an  super-nova driven outflow. 
 If it had evolved as closed box instead (untill the same gas fraction), the amount
 of metals it would have is given by
 \begin{eqnarray}
 M_Z^{\rm cb}&=&Z^{\rm cb} (M_{\rm ISM}^{\rm cb}+M^{\rm cb}_{\rm stars}) ,\nn\\
 &=& y_0 \ln(1/\mu) (M_{\rm ISM}^{\rm cb}+M^{\rm cb}_{\rm stars}),
 \end{eqnarray}
where $M_{\rm ISM}^{\rm cb}+M_{\rm stars}^{\rm cb}$ is the baryonic mass $M_{\rm bar}^{\rm cb}$ and $y_0$ is the true yield.

Instead, it is observed to be metal poor for its gas fraction $\mu$. 
The amount of stars $M_{\rm stars}$ with or without outflow would be the same
as it depends only on the star formation rate.
In the case of outflows made purely of SN ejecta,
the amount of gas $M_{\rm ISM}$ left after the outflow episode is also
the same.
This SN ejecta simply left the ISM and failed to enrich the remaining gas and stars.
If the outflow entrained some fraction of the ISM, the gas mass 
may differ from that of the closed box evolution.

In the most general case, the total amount of metals after an outflow
episode is:
 \begin{eqnarray}
 M_Z^{\rm w}&=&Z^{\rm w} (M_{\rm ISM}^{\rm w}+M^{\rm w}_{\rm stars}) ,\nn\\
 &=& y_{\rm eff} \ln(1/\mu) (M_{\rm ISM}^{\rm w}+M^{\rm w}_{\rm stars}),
 \end{eqnarray}
where $y_{\rm eff}$ is the effective yield.

The amount of metals that were lost is $M_Z^{\rm lost}=M_{Z}^{\rm cb}-M_{Z}^{\rm w}$.
The fraction of metals lost relative to what is left $M_Z^{\rm w}$ is 
\begin{eqnarray}
\frac{M_Z^{\rm cb}-M_Z^{\rm w}}{M_Z^{\rm w}}
&=& \frac{M_Z^{\rm cb}}{M_Z^{\rm w}}-1\nn\\
&=& \frac{y_0}{y_{\rm eff}}\;\frac{M_{\rm bar}^{\rm cb}}{M_{\rm bar}^{\rm w}} - 1 \label{eq:appendix:yeff}\\
&\simeq& \frac{y_0}{y_{\rm eff}} - 1 \label{eq:appendix:yeff:approx}
\end{eqnarray}
Since the baryonic mass is not changed by outflows made of 100\%\ ejecta,
and is likely to be close to its original baryonic mass even in the case
of outflows made of entrained material, Eq.~\ref{eq:appendix:yeff}
gives the ratio between the amount of metals lost from galaxies
to that the amount of metals left.

\subsection{Dependence on the parameters}
 \label{appendix:dependence}

The results presented in Figures~\ref{fig:garnett:z0}--\ref{fig:ejected:z3},
 can be understood from Eqs.~\ref{eq:Garnett:Mlos} and  \ref{eq:masslost:oxygen}.
The core ingredient  is the effective yield, $y_{\rm eff}$.
Because $y_{\rm eff}$ increases with mass or luminosity 
the term $\frac{y_0}{y_{\rm eff}}-1$ decreases with $L_B$.
In the case of Eq.~\ref{eq:Garnett:yeff} (as it can be estimated from Fig.~\ref{fig:effective_yield})
the effective yield  goes as $y_{\rm eff}\propto L_B^{+1/3}$.
In the case of Tremonti's parameterization (Eq.~\ref{eq:Tremonti:yeff}), 
$y_{\rm eff}\propto V_{\rm rot}^2 \propto L^{+2/3}$
using $V_{\rm rot}\propto L^3$ from the $B$-band Tully-Fischer relation.
In what follows, we parameterize $y_{\rm eff}\propto L_B^{\beta}$.
Because the LZ relation increases with $L_B$ \citep[e.g][]{LequeuxJ_79a,SkillmanE_89a,ZaritskiD_94a,PilyuginL_04a},    the factor 
$\frac{\rm O}{\rm H} L_B$ in Eq.~\ref{eq:Garnett:Mlos}
is a  steep function of $L_B$, it is $\propto L^{1+2.5\times0.16}_B\propto L^{1.4}_B$.
As a consequence, the amount of metals lost per galaxy according to Eq.~\ref{eq:Garnett:Mlos}
goes as $M_{\rm O,\rm lost} \propto L_B^{-\beta}\;L_B^{1.4 }$:
\begin{eqnarray}
M_{\rm O,\rm lost}(L_B)&\propto& L_B^{+0.73} \hbox{for $\beta=2/3$},\\
M_{\rm O,\rm lost}(L_B)&\propto& L_B^{+1.06} \hbox{for $\beta=1/3$},
\end{eqnarray}
which means that a bigger galaxy ejects more metals than a dwarf in absolute terms.

The shape of the comoving density of metals per unit luminosity (given by Eq.~\ref{eq:masslost:oxygen})
 can now be estimated since it is given by 
the amount of metals per galaxy times the luminosity function.
 At the faint end of the LF, $\alpha\simeq -1.0$, such
that the comoving density of metals per unit luminosity 
 will be almost constant:
\begin{eqnarray}
\frac{\mathrm d M_{\rm O}}{\mathrm d V\mathrm d L} & \propto &  L_B^{-0.27}  \hbox{for $\beta=2/3$},\\
\frac{\mathrm d M_{\rm O}}{\mathrm d V\mathrm d L} & \propto &  L_B^{+0.06}  \hbox{for $\beta=1/3$}.
\end{eqnarray}
and the comoving density of metals per unit magnitude 
will increase steeply:
\begin{eqnarray}
\frac{\mathrm d M_{\rm O}}{\mathrm d V\mathrm d {\rm mag}} & \propto & L_B^{\sim 0.7} \hbox{for $\beta=2/3$},\\
\frac{\mathrm d M_{\rm O}}{\mathrm d V\mathrm d {\rm mag}} & \propto & L_B^{\sim 1.0} \hbox{for $\beta=2/3$}.
\end{eqnarray}
The steep exponential cut off of the bright end of the LF will make the contribution of
large galaxies to the comoving density of metals per unit magnitude drop sharply. 
Thus, somewhere between  $L<<L^*$ and $L>>L^*$, the distribution  $\frac{\mathrm d M_{\rm O}}{\mathrm d V\mathrm d mag}$
(Eq.~\ref{eq:masslost:oxygen}) will peak at an absolute magnitude $M_B\la L^*$.
This conclusions
does not change  {\it even if} $M/L_B$ depends slightly on $L_B$.

\section{Assumptions and ingredients}

In this Appendix, we detail the ingredients used in our calculations
of the global amount of metals ejected from galaxies
presented in sections~3.2, and 3.3.

\subsection{Metals lost at $z=0$}
\label{appendix:z0}

\begin{table}
\caption{Parameters used in section~\ref{section:ejected:z0} and Fig.~\ref{fig:garnett:z0}.\label{table:LB}}
\begin{tabular}{lrrrr}
\hline
Parameter 	& Value	&	 Error	&	Relative  	&	Error	  \\
		&	&	&	Error  		&  	to $\rho_{\rm O}$  \\
		&	&	(1-$\sigma$)	&	 (\%)		&  (1-$\sigma$) \\
	\hline 
$y_{\rm eff}$	&	&		&	25	&  $^{+1.3\hbox{E}6}_{-1.1\hbox{E}6}$ \\
LF $\phi_\star$ & 0.0088&	0.002	&		&  $^{+9.1\hbox{E}5}_{-9.1\hbox{E}5}$ \\
LF $M_\star$	& -20.27&	0.1	&		&  $^{+1.7\hbox{E}5}_{-1.7\hbox{E}5}$ \\
LF $\alpha$	& -1.05	&	0.05	&		&  $^{+3.0\hbox{E}5}_{-2.7\hbox{E}5}$ \\
TF slope	&  -7.3	&	0.5	&		&  $^{+3.4\hbox{E}5}_{-2.8\hbox{E}5}$ \\
TF intercept	& -20.11&	0.15	&		&  $^{+5.0\hbox{E}5}_{-4.6\hbox{E}5}$\\
LZ slope	& -0.16	&	0.01	&		&  $^{+2.0\hbox{E}6}_{-1.3\hbox{E}6}$ \\
LZ intercept	& -6.4	&	0.10	&		&  $^{+9.9\hbox{E}5}_{-7.9\hbox{E}5}$ \\
$M/L$		& 6	&	2	&		&  $^{+1.3\hbox{E}6}_{-1.3\hbox{E}6}$\\
\hline								   	  	   
Uncertainty to Eq.~\ref{eq:Mlos:z0} &	& 	& 	&  $^{+3.1\hbox{E}6}_{-2.5\hbox{E}6}$\\
\hline
\end{tabular}
\end{table}

Here, we list the ingredient that went in Eq.~\ref{eq:Garnett:Mlos}.
First, for the $y_{\rm eff}$  as a function of 
  rotational velocity $V_{\rm rot}$,
   \citet{GarnettD_02a}  parameterized it as  (for galaxies with $V_{\rm rot}<250$~\kms)
\begin{equation}
\log y_{\rm eff}=\log {y_0}{-\frac{(320-V_{\rm rot})^4}{0.1\times 10^9}} \label{eq:Garnett:yeff}
\end{equation}
where $y_0=10^{-1.95}=0.01122$ is the true yield for Oxygen (O).

The second central ingredient in Eq.~\ref{eq:Garnett:Mlos} is the well known
 \LZ\ relation.
At  $z=0$, \citet{GarnettD_02a} characterized  it as:
\begin{equation}
12+\log \frac{\rm O}{\rm H}=-0.16 \; M_B -6.4 \label{eq:Garnett:LZ}\;.
\end{equation}
Similarly, \citet{TremontiC_04a} found $12+\log \frac{\rm O}{\rm H}=-0.18 \; M_B -5.238\;. $\

The third important ingredient is the LF and the $B$-band TF-relation. We used 
a LF with $\alpha=-1.05$, $\phi_*=0.017*h^3$ and $M^*-5\log h=-19.65$ from \citet{CrotonD_05a} similar
to that of \citet{LiskeJ_03b},
and the $B$-band Tully-Fischer \citep[e.g.][]{TullyB_00a}:
%Vrot=0.5*(10**(2.5+(M_B+20.11)/(-7.3)))
$M_B=-7.3\;\log[ 2 V_{\rm rot} -2.5]-20.11$.

The final and critical ingredient is the 
value of  $M/L_B$~\footnote{It should be noted that $M$ in the mass-to-light ratio $M/L_B$ should be the baryonic mass since 
the factor $y_0/y_{\rm eff}-1$ estimates the fraction of
ejected metals, as a fraction of the total baryonic mass of a galaxy.}
as it sets the normalization in computing $\rho_{Z,\rm ejected}$ (Eq.~\ref{eq:masslost:oxygen}).
Many have estimated the mean stellar mass-to-light ratios
\citep[e.g.][]{BellE_01a,KauffmannG_03b}.
The M/L ratio varies strongly as a function of $B$-$V$ color \citep{BellE_01a,RudnickG_06a}, or galaxy type \citep{ReadJ_05a},
from 1 to 9.
One way to estimate $\langle M/L_B\rangle$ is from
the ratio of the stellar density $\rho_\star$ to $B$-band luminosity density $j_B$.
There has been many estimates of  the stellar mass function at $z=0$ using various methods.
 For instance, \citet{ColeS_01a} computed the global stellar density $\rho_\star$ from the
 $K$-band luminosity function of 2dF galaxies, and found $\rho_\star=5.6\;\times10^{8}$~\rhosun.
\citet{BellE_03a} used a similar approach (using a 'diet' Salpeter IMF)
and found that $\rho_\star=3.7 \times 10^8 h_{70}$~\rhosun, corresponding
$\rho_\star=5.9 \times 10^8 h_{70}$~\rhosun\ for a Salpeter IMF, very similar
to the results of \citet{ColeS_01a}. 
Recent estimates by  \citet{ReadJ_05a} showed that
 $\Omega_\star=0.0028$ ($\rho_\star=3.78\;10^{8}$~\rhosun) using a Kroupa IMF,
 which corresponds to $\rho_\star=7.5\;10^{8}$~\rhosun\ for a Salpeter IMF.
 \citet{PanterB_04a} found 
   for a Salpeter IMF  $\Omega_\star=0.0034$ ($\rho_\star=4.6\times10^8$~\rhosun). 
We will use $$\rho_\star=6\times10^8~\rhosun.$$
Using 2dF, \citet{MadgwickD_02a} and
\citet{CrotonD_05a} found that $$j_{B_J}=1.4\times10^8\;h^1_{70}~\Lsun.$$
% According to \citet{RudnickG_03a}, 
%$j_B(z=0)=7.4\;10^8$~$L_\odot$~Mpc$^{-3}$, implying that
%the mean $\langle M_\star/L_B \rangle$ is about $\sim 8$.\
Thus, the mean  $\langle M_\star/L_B \rangle$ is   $\sim 4$.
Given that 80\%\ of the baryons in galaxies are in stars \citep{ReadJ_05a},
a reasonable baryonic mass-to-light ratio is $\langle M_{\rm bar}/L_B\rangle\;\sim\; 6$.
 
	Table~\ref{table:LB} shows the parameters used along  with their contribution to the error budget.

\subsection{Metals lost at $z=0$ in terms of stellar masses}
\label{appendix:stellarmass}

In terms of stellar masses, Eq.~\ref{eq:Garnett:Mlos} then becomes in terms of $M_\star$ :
\begin{equation}
M_{O, \rm lost}(M_\star)=12\frac{\rm O}{\rm H}(M_\star) M_\star \left [\frac{y_0}{y_{\rm eff}(M_\star)}-1\right
]\msun \;. \label{eq:Garnett:Mlos2}
\end{equation}

The relationships needed here are  the stellar mass-metallicity ($M_\star$--$Z$),
the $y_{\rm eff}(M_\star)$, the stellar TFR, and the stellar mass function.
  \citet{TremontiC_04a} 
constructed the $M_\star$--$Z$  relation from 53,000 SDSS galaxies
and found:
\begin{equation}
12 + \log \frac{\rm O}{\rm H}=-1.49 + 1.85\;(\log M_\star) -0.08 \; (\log M_\star)^2\;. \label{eq:Tremonti:LZ}
\end{equation}
 They also modeled   the effective yield as 
 \begin{eqnarray}
\log \frac{y_{\rm eff}}{y_o}&=&-\log\left [1+\left (\frac{V_o}{V_{\rm rot}}\right )^2 \right ]\label{eq:Tremonti:yeff}
 \end{eqnarray}
 where $y_0=0.0104$ and $V_0=85$~\kms.
% Since, the Baryonic TF relation is $M\propto V_{\rm rot}^3$, $y_{\rm eff}\propto L_B^{-2/3}$.
We used the stellar TF relation $M_\star$--$V_{\rm rot}$ of \citet{BellE_01a}:
\begin{equation}
\frac{M_\star}{10^{9.8}\msun}=\left(\frac{V}{100\kms}\right)^{3.5} \;.\label{eq:TF:bellE}
\end{equation}
The final ingredient is the stellar mass function from  \citet{ReadJ_05a}.

\subsection{Metals lost at $z=2.5$}
\label{section:appendix:z2}

For our calculation at $z=2.5$ in section~\ref{section:ejected:z2},
we used the same  $y_{\rm eff}$ relation from \citet{GarnettD_02a} (Eq.~\ref{eq:Garnett:yeff}).
Whether the $y_{\rm eff}$--$V_{\rm rot}$ relation is the same at those redshifts is still an open question.
However, one can argue that this is  likely  for the following reasons: (i) the $z=0$
  $y_{\rm eff}$--$V_{\rm rot}$ relation is made of galaxies of very different
 star formation histories (i.e. forming over a wide range of redshifts),   (ii) $V_{\rm rot}$ is a measure of 
 the potential well, which is a key factor determining the outcome of the galactic super-wind phenomenon
  \citep[e.g.][]{AguirreA_01a}, and (iii) both local \citep[e.g.][]{LehnertM_96a,HeckmanT_00a,DahlemM_97a}  and high-redshift starbursts 
  \citep[e.g.][]{PettiniM_01a,AdelbergerK_03a} drive strong galactic winds.

We note that the parameterization of $y_{\rm eff}$--$V_{\rm rot}$ relation by
\citet{TremontiC_04a} (Eq.~\ref{eq:Tremonti:yeff}) diverge at the low mass end, and  
our result would  increased by 25\%, a factor that  depends
strongly on the minimum luminosity $M_{\rm min}$ used in  the integration of $\frac{\mathrm d M_{\rm O}}{\mathrm d V\mathrm d mag}$.
 We used the parameterization of $y_{\rm eff}$--$V_{\rm rot}$ relation of
\citet{GarnettD_02a} (Eq.~\ref{eq:Garnett:yeff}), which 
does not diverges and thus produces results that do not
depend on the minimum luminosity $M_{\rm min}$.

For the LF, \citet{SteidelC_99a} constrained   
 the faint end slope of the $z=3$ LF and found it
to be much steeper $\alpha=-1.60$  than the local LF.
However,   recent detailed work by    \citet{SawickiM_06a}
did not confirm such a steep LF.
They  found $\alpha=-1.43\pm0.15$
  at $z=3$ and $\alpha=-1.2\pm0.2$ at $z=2$.
\citet{GiallongoE_05a} studied the evolution of the LF in a parametric way,
and kept the faint end fixed to $\alpha=-1.3$ with redshift.
They found that $M^*_{4400\AA,\rm AB}(z=2.2)=-20.93-1.48\;\log(1+2.2)$ or
$M^*_{4400\AA, \rm AB}=-21.67$. 
\citet{SawickiM_06a} found that $M^*_{1700,\rm AB}=-20.60$, which
corresponds to $M^*_{4400,\rm AB}=-21.60$ where the $K$-correction was $-1.03$
assuming a SED slope of $\beta=-1$.
Both studies   found similar $\phi^*$: $\phi^*=0.0026$ ($\phi^*=0.003$)
according to \citet{GiallongoE_05a} \citep{SawickiM_06a}. 
We converted the numbers for a $h=0.7$ $\Lambda$CDM universe when necessary.

With regards to the mass-metallicity or luminosity relation, 
 \citet{KobulnickyH_00a}, \citet{PettiniM_01a} 
and \citet{MehlertD_02a}  have shown that  $z\sim 3$  galaxies
 are 2--4 mag over-luminous for their metallicities, 
 when compared with the local metallicity-luminosity relation. 
 A conclusion also reached by \citet{ShapleyA_04a} at $z\sim2$ on a small sample of 8 galaxies.
Conversely, at a given luminosity, galaxies are a factor 2 to 3 more metal poor. 
 Following on this work, \citet{ErbD_06a} used a much larger sample of 87 $z\sim2.2$ galaxies
and constrained the stellar mass--metallicity ($M_\star$--$Z$)    relation.
They found that $z\sim2.2$ UV-selected galaxies are 0.3dex  less metal rich at a given stellar mass
than the $z=0$ sample of \citet{TremontiC_04a}~\footnote{\citet{ErbD_06a} converted the \citet{TremontiC_04a} metallicities to the same  
[NII/\Ha] scale.}.  \citet{ErbD_06a} also found that the 
scatter in the $M_\star$--$Z$ relation is much less than in the $L_B$--$Z$ relation, owing to the 
larger scatter in mass-to-light $M/L_B$ ratios.

The evolution of the TFR to $z\sim 2$ (10~Gyr ago) is unknown, and its evolution to $z=1$
is  debated. Whether it evolves or not appears to depend mostly on the photometric band.
In  summary, the $B$-band TFR seems to be offsetted from the local TFR by about 1~mag
 \citep{Milvang-JensenB_03a,BoehmA_04a,NakamuraO_06a,BamfordS_06a,WeinerB_06a}.
However, \citet{VogtN_01a}   argued that most of the evolution is due to selection effects.
The $K$-band TFR evolution appears negligible \citep{ConseliceC_05b,FloresH_06a}.
Given the uncertain conclusions,  we will
 first assume an offset of 1~mag for the  $z\sim 2$ TFR, and repeat the calculation with no evolution of the TFR.

The mass-to-light ratio $M/L_B$ is much different at $z=2$--3 than at $z=0$. \citet{BellE_01a} 
showed that $M/L_B$ is a strong function of color and is about 1.5 for the bluest 
(youngest) galaxies. Similarly to the previous $z=0$ calculation, we can take
the ratio between $j_B$ and $\rho_\star$ at $z\sim2.2$ to estimate the
mean $\langle M/L_B\rangle$ ratio.
\citet{RudnickG_03a} and our Fig.~\ref{fig:SFH} showed that $\rho_\star(z=2.2)\simeq 2\times10^{8}$~\rhosun.
\citet{RudnickG_03a} estimated $j_B$(rest) at $z=2.2$ and found that $j_B\simeq10^{8\hbox{--}8.3}$.
Thus, $\langle M/L_B\rangle$ is about 1 to 2.
 \citet{ErbD_06a} showed that the $M/L_B$ ratio
  for UV-selected galaxies at $z=2$, ranges from 0.02 to 1.4. 
We will use a stellar mass-to-light ratio of 1.
 \citet{ErbD_06a} also  indicated that these galaxies
 have large gas factions, from 20\%\ up to 80\%, a result in contrast to local galaxies
 where the gas fractions are $\sim15$\%\ on average (see Table~\ref{table:bigsummary}).
 Thus, a reasonable baryonic mass-to-light ratio for high redshift galaxies is 2.

Table~\ref{table:error:z3} shows the error analysis for each of the parameters coming into
play in our calculation. One sees that the dominant uncertainties comes from   
$y_{\rm eff}$ and the mean $\langle M/L\rangle$ ratio and
our result (Eq.~\ref{eq:Mlos:z2}) is uncertain by a factor of $\sim2$.

\begin{table}
\caption{Parameters used in section~\ref{section:ejected:z2} .\label{table:error:z3}}
\begin{tabular}{lrrrr}
\hline
Parameter 	& Value	&	 Error	&	Relative  	&	Relative	  \\
		&	&	(1-$\sigma$)&	Error  		& Error	to $\rho_{\rm O}$  \\
		&	&		&	 (\%)		&  (\%) \\
	\hline 
$y_{\rm eff}$ & 	&		& 25	&  $^{+1.1\hbox{E}6}_{-1.2\hbox{E}6}$ \\
LF $\phi_\star$ & 0.003	& 0.001		& 30	&  $^{+7.4\hbox{E}5}_{-7.4\hbox{E}5}$ \\
LF $M_\star$	& -21.69& 0.25  	& 1.1 	&  $^{+2.9\hbox{E}5}_{-2.7\hbox{E}5}$ \\
LF $\alpha$	& -1.3	& 0.2		& 15	&  $^{+1.4\hbox{E}6}_{-7.3\hbox{E}5}$ \\
TF slope	& -7.3	& 0.5		& 7	&  $^{+1.4\hbox{E}5}_{-1.2\hbox{E}5}$	\\
TF intercept	& -20.11& 0.25		& 1	&  $^{+4.0\hbox{E}5}_{-3.5\hbox{E}5}$ \\
TF offset	& -1    & 0.25		& 25	&  $^{+4.0\hbox{E}5}_{-3.5\hbox{E}5}$ \\
LZ slope	& -0.185& 0.01		& 5	&  $^{+1.2\hbox{E}6}_{-8.0\hbox{E}5}$  \\
LZ intercept	& 5.238  & 0.2  	& 4	&  $^{+1.3\hbox{E}6}_{-8.2\hbox{E}5}$\\
LZ offset	&  $-$0.30 &  	0.07	& 25  	&  $^{+4.5\hbox{E}5}_{-2.8\hbox{E}5}$\\
$M/L$		&  2	&  0.67		&   30	&  $^{+1.1\hbox{E}6}_{-1.1\hbox{E}6}$	\\
\hline
Uncertainty to Eq.~\ref{eq:Mlos:z2}
	&	& 	&	& $^{+2.9\hbox{E}6}_{-2.2\hbox{E}6}$\\
\hline
\end{tabular}
\end{table}

\section{Metal Budget}

In this Appendix, Table~\ref{table:bigsummary} summarizes the metal budget.
Some classes of objects are repeated and care should be taken in combining different entries.

%%%%%%%%%%%BIG TABLE
%%% FROM Baryon_budget.txt
%% with
%% awk 'BEGIN {FS=","; OFS="\t& "} sub(/[eE]/,"times10^{",$2)!=0  {print $1," $"$2"}\\;"$3"$ "," $"$4"\\;"$5"$ ",$6,$7,$8,"$"gensub(/[eE]/,"times10^{",1,$10)"}$","$"gensub(/[eE]/,"times10^{",1,$11)"}$", $12, $13 " \\\\" } ' Baryon_budget.txt | sed 's/times/\\times/g' | sed 's/\$}\$/ /g' | sed 's/\$ }\$/ /g' | sed 's/h/h_{70}/g' | sed 's/+0//g' | sed 's/{+/{/g' | sed 's/-0/-/g' > Baryon_budget.tex
%% awk 'BEGIN {FS=","; OFS="\t& "} sub(/[eE]/,"times10^{",$2)!=0  {print $1," $"$2"}$ "," $"$4"$ ",$6,$7,$8,"$"gensub(/[eE]/,"times10^{",1,$10)"}$","$"gensub(/[eE]/,"times10^{",1,$11)"}$", $12, $13 " \\\\" } ' Baryon_budget.txt | sed 's/times/\\times/g' | sed 's/\$}\$/ /g' | sed 's/\$ }\$//g' | sed 's/h/h_{70}/g' | sed 's/+0//g' | sed 's/{+/{/g' | sed 's/-0/-/g' > Baryon_budget.tex

\setlength{\tabcolsep}{1pt}

\begin{table*}
%\begin{deluxetable}{lclrl|ccr}
\caption{Contributions of various classes of objects. We use $Z_\odot=0.0189$, and 
in our cosmology, the critical density is $\rho_c=1.36\times10^{11}\;h_{70}^{2}\rhosun$.
}
\begin{tabular}{lclrlccccrr}
\hline
		&  $\rho $ 	& $\rho/\rho_c$ 	 	&  $\rho/\rho_b$  	 & Refs~\tablenotemark{a} &	$Z/Z_\odot$ &
		$\rho_Z$  	& $\rho_Z/\rho_c$  & $\overline Z_b$	&   $\rho_Z/\rho_{Z,\rm tot}$ & Refs~\tablenotemark{a}	\\
		&  $ (\rhosun)$ 	&  	 	&   (\%) 	 &   &	  &
		 (\rhosun) 	&    & ($Z_\odot$)	&  (\%) &  	\\
\hline  
Baryons ($\rho_b$)	&  $ 5.98\times10^9$ &  $0.044$	&   100	& 1,2    	 \\
\hline\hline
\multicolumn{10}{c}{$z=0$}   \\
\hline\hline
\multicolumn{3}{l}{All Metals}	& &  		&	& $7.75\times10^{7}$	& $5.70\times10^{-4}$	&  0.6854 &	&  4\\
MS stars	&	&	& &  		&	& $4.35\times10^{6}$    & $3.20\times10^{-5}$   &  0.0385 &       &  4\\
WDs(C+O)	&  	&	& &		&       & $4.90\times10^{7}$	& $3.60\times10^{-4}$	& 0.4329 & 	 & 4 \\
\hline
\multicolumn{3}{l}{Metals produced by Type~II SNs}	  &	& 
		&  	& $2.30\times10^{7}$	& $1.69\times10^{-4}$	& 0.2034	& 100.0  &    \\\hline
Stars $ z>0$	&  $6.43\times10^{8}$ &  $0.004728$ 	& 10.75	& $\int$ SFR	& 1/42~\tablenotemark{b}	
& $2.13\times10^{7}$	& $1.56\times10^{-4}$	& 	&	 	&  \\
Stars	
	&  $5.63\times10^{8}$ 	&  $0.004140$ 	& 9.41	& 1,2,5  (Salp.)	& 1	& $1.48\times10^{7}$	& $1.09\times10^{-4}$	& 0.1307	& 64.3	&  \\
%	&  $6.53\times10^{8}$ 	&  $0.004801$ 	& 10.91	& 1 (S IMF)	& 	&  	&  	& 	& 	&  \\
%	&  $4.94\times10^{8}$ 	&  $0.003632$ 	& 8.26	& 2 (S IMF)	& 	&  	&  	& 	& 	&  \\
\hline
BH+NS 		&   	&  	& &  		& 	& $1.60\times10^{7}$  	& $1.18\times10^{-4}$   & 0.1419 & 69.9	& 4    \\  
\hline
$\HI$	&  $5.71\times10^{7}$	      &  $0.000420$	 & 0.95  & 8, 9  &	 &	 &	 &	 &  \\
$\Htwo$	&  $2.17\times10^{7}$	      &  $0.000160$	 & 0.36  & 10	 &	 &	 &	 &	 &  \\
\hline
Cold gas ($+\HeI$)
	&  $1.06\times10^{8}$ 	&  $0.000782$ 	& 1.78	& (8+10)$\times$1.35	& 0.84	& $1.69\times10^{6}$	& $1.24\times10^{-5}$	& 0.0149	& 7.3	&  \\
\hline
Hot gas (cl.)	&  $3.57\times10^{8}$ 	&  $0.002625$ 	& 5.97	& 3	& 	&  	&  	& 	& 	&  \\
	&  $2.45\times10^{8}$ 	&  $0.001801$ 	& 4.09	& 4	& 0.3333	& $1.54\times10^{6}$	& $1.13\times10^{-5}$	& 0.0136	& 6.7	& 10 \\
\hline
Hot gas(gr.)	&  $6.80\times10^{8}$ 	&  $0.005000$ 	& 11.36	& 3	& 0.3333& $4.28\times10^{6}$	& $3.15\times10^{-5}$	& 0.0379 	&18.6	&   \\
\hline
WHIM  IGM OVI	&  $>3.67\times10^{8}$ 	&  $0.002700$ 	& 6.14	& 11	& 0.1?	& $6.94\times10^{5}$	& $5.10\times10^{-6}$	& 0.0061	& 3.0	&  \\
WHIM  IGM BLyAb	&  $<7.89\times10^{8}$ 	&  $0.005800$ 	& 13.18	& 11,18	& 0.1?	& $1.49\times10^{6}$	& $1.10\times10^{-5}$	& 0.0132	& 6.5	&  \\
\hline\hline
\multicolumn{10}{c}{$z=2.5$} \\
\hline\hline
Stars $ 2<z<4$	&  $1.21\times10^{8}$ 	&  $0.000890$ 	& 2.02	& $\int$ SFR	& 1.260	& $4.00\times10^{6}$	& $2.94\times10^{-5}$	& 0.0354	& 100.0	&  \\
\hline
%LBGs	&  $4.23\times10^{7}$ 	&  $0.000311$ 	& 0.71	& 17	& 0.3	& $2.40\times10^{5}$	& $1.76\times10^{-6}$	& 0.0021	& 6.0	&  \\
%	&  $2.19\times10^{7}$ 	&  $0.000161$ 	& 0.37	& 12	& 0.3	& $1.24\times10^{5}$	& $9.13\times10^{-7}$	& 0.0011	& 3.1	&  \\
BX	&  $4.02\times10^{7}$ 	&  $0.000185$ 	& 0.67	& 	& 0.5	& $3.80\times10^{5}$	& $2.79\times10^{-6}$	& 0.0034	& 9.5	&  \\
BX+K20	&  $1.98\times10^{7}$ 	&  $0.000146$ 	& 0.33	& 	& 0.8	& $3.00\times10^{5}$	& $2.21\times10^{-6}$	& 0.0027	& 7.5	&  \\
DRG	&  $1.06\times10^{7}$ 	&  $0.000078$ 	& 0.18	& 	& 1	& $2.00\times10^{5}$	& $1.47\times10^{-6}$	& 0.0018	& 5.0	&  \\
SMGs $>3$mJy&  $2.73\times10^{7}$ &  $0.000201$ & 0.46	&  	& $<$1.86	& $<$$3.63\times10^{5}$	& $<$$2.67\times10^{-6}$	& $<$0.0032
 & $<$9.1	&  \\
\hline
DLA ($N>20.3$)	&  $1.50\times10^{8}$ 	&  $0.001103$ 	& 2.51	& 13	& 0.07	& $1.98\times10^{5}$	& $1.46\times10^{-6}$	& 0.0018	& 5.0	&  \\
30\%\ missed	&  $5.00\times10^{7}$ 	&  $0.000368$ 	& 0.84	& 14	& 0.7	& $6.62\times10^{5}$	& $4.86\times10^{-6}$	& 0.0058	& 16.5	&  \\
\hline
LLS $19$--$20.3$&  $3.60\times10^{7}$   &  $0.000260$   & 0.60 	& 19	& 0.1   & $<6.80\times10^{5}$    & $<5.00\times10^{-6}$   & $<$0.0060        & $<$17.0  & 19\\
LLS $19$--$20.3$&  $2.72\times10^{7}$   &  $0.000200$   & 0.45 	& 19    & 0.15  & $>7.71\times10^{4}$    & $>5.67\times10^{-7}$   & $>$0.0007        & $>$2.0   & 19\\ 
LLS $17$--$19$ & ...&...&...&...&...&...&...&...&...&  \\
\hline
Forest	&  $5.44\times10^{9}$ 	&  $0.040000$ 	& 90.91	& 3	& 0.003	& $3.08\times10^{5}$	& $2.27\times10^{-6}$	& 0.0027	& 7.7	&  \\
Forest	&  $6.14\times10^{9}$ 	&  $0.045130$ 	& 102.6	& 	& $<$0.005	& $<5.80\times10^{5}$	& $<4.26\times10^{-6}$	& $<$0.0051	& $<$14.5	& 19 \\
\hline

%\enddata
%\end{center}
\tablenotetext{a}{References:   (1) \citet{KochanekC_01a}, (2) \citet{BellE_03a}, (3) \citet{FukugitaM_98a}, (4) \citet{FukugitaM_04a}, (5) \citet{ColeS_01a}, (6) \citet{BellE_03a},
(7) \citet{RudnickG_03a}, (8) \citet{ZwaanM_03a}, (9) \citet{RosenbergJ_03a}, (10) \citet{KeresD_03a}, (11) \citet{SembachK_04a}, (12) \citet{DunneL_03a}, (13) \citet{PerouxC_03a},
(14) \citet{VladiloG_04a}, (15) \citet{SimcoeR_04a}, (16) \citet{SchayeJ_01a}, (17) \citet{PettiniM_03a}, 
(18) \citet{RichterP_06a}, (19) This paper.
}
\tablenotetext{b}{Averaged yield for type II SN with $m>10$~\msun\ \citep{MadauP_96a}, taking into account 
a recycled fraction fo $R=0.28$.}
\end{tabular}
\label{table:bigsummary}
%\end{deluxetable}
\end{table*}

\bsp_small

\label{lastpage}

\end{document}